\documentclass[12pt]{article}
\usepackage{latexsym,amssymb,amstext,amsmath,array}
\usepackage{amscd}
\usepackage{cite}
\usepackage{slashed}
\usepackage{mathrsfs}
\usepackage{hyperref}
\usepackage{xcolor}

\newcommand {\cA}{{\cal A}}

\newcommand {\cC}{{\cal C}}

\newcommand {\cJ}{{\cal J}}

\newcommand {\cL}{{\cal L}}
\newcommand {\cM}{{\cal M}}
\newcommand {\cN}{{\cal N}}

\newcommand {\cQ}{{\cal Q}}
\newcommand {\cR}{{\cal R}}
\newcommand {\cS}{{\cal S}}

\newcommand {\cY}{{\cal Y}}

\newcommand{\bR}{{\bf R}}

\def\a{\alpha}
\def\b{\beta}

\def\d{\delta}
\def\e{\epsilon}

\def\g{\gamma}
\def\G{\Gamma}

\def\l{\lambda}

\def\o{\omega}

\def\z{\zeta}

\def\L{\Lambda}

\def\U{\Upsilon}

\newcommand{\cbar}{\bar{c}}

\newcommand{\be}{\begin{equation}}
\newcommand{\ee}{\end{equation}}
\newcommand{\bea}{\begin{eqnarray}}
\newcommand{\eea}{\end{eqnarray}}

\newcommand{\ba}{\begin{array}}
\newcommand{\ea}{\end{array}}

\def\double #1{#1{\hbox{\kern-2pt $#1$}}}

\newcommand{\bsubeq}{\begin{subequations}}
\newcommand{\esubeq}{\end{subequations}}

\begin{document}

\begin{titlepage}

\begin{center}

\vskip .3in \noindent

{\Large \bf{$N=4$ Supersymmetric BMS$_3$ algebras\\[1mm] from asymptotic symmetry analysis %% BMS$_3$ to all BPS
}}

\bigskip

	{Nabamita Banerjee}$^{\,a}$
	\footnote{nabamita@iiserpune.ac.in}, {Ivano Lodato}$^{\,a}$\footnote{ivano@iiserpune.ac.in}, {Turmoli Neogi}$^{\,a}$\footnotetext{turmoli.neogi@students.iiserpune.ac.in}\\

       \bigskip
       $^{a}$ \em Indian Institute of Science Education and Research,\\ Homi Bhabha Road, Pashan, Pune 411 008, India \\

       \vskip .5in
       {\bf Abstract }
       \vskip .2in
       \end{center}
        We consider three dimensional $N=4$ flat supergravity, with an 
        abelian R-symmetry             		enhancing the gravitational phase space. We obtain the field configuration whose 				asymptotic symmetries at null infinity coincide with the centrally extended $N=4$ 				super Bondi-Metzner-Sachs (BMS) algebra. The killing spinors for this generic configuration are obtained 				together with the energy bounds imposed by  supersymmetry. It is explicitly shown that 			the same algebra can be obtained  as a flat (AdS radius $\rightarrow \infty$ ) limit of 		the combined $(2,0)$ and $(0,2)$ sectors of AdS supergravity.

       %\vskip .4in

\vfill
\eject

\end{titlepage}
\newpage
\tableofcontents
\section{Introduction and Summary}

 Supergravity theories in $2+1$ dimensions have many interesting features which have no equivalent in their higher-dimensional counterparts. It is well-known 
 for instance, that no local degrees of freedom exist in the bulk and that 
 it is not possible to define the linear momentum or the supercharges for any  
 solution at spatial infinity \cite{Ashtekar, PhysRevD.29.2766, 0264-9381-1-1-001}. One can only define the energy and the 
 angular momentum there associated to asymptotic time translations and spatial rotations. However, the scenario changes at the null infinity.
  Almost half a century ago, in their seminal works \cite{Bondi,Sachs},  Bondi, 
 van der Burg, Metzner and independently Sachs first introduced the symmetries 
 of 4D flat space times at their null infinity, named BMS symmetry. Later, in 
 \cite{Ashtekar,Barnich, BarnichT}, it has been shown that the asymptotic structure for flat three-dimensional 
 gravity at their null infinity is also much richer: it consists of an infinite 
 dimensional symmetry whose generators, super-translation and super-rotation generators, act on the boundary coordinates.\\
A similar symmetry enhancement also takes place when one considers the asymptotic 
algebra of symmetries of three dimensional AdS (super)gravity. In their  seminal
paper \cite{BrownHenneaux}, Brown and Henneaux showed that, upon imposing suitable 
boundary conditions for the fields in asymptotically AdS$_3$ space, the symmetry 
enhances from SO$(2,2)$ to the infinite dimensional conformal algebra in two 
dimensions. This is connected to the enhancement of the flat asymptotic algebra,
as the latter corresponds to a well defined flat space limit of the AdS algebra  
\cite{Fareghbal1,Barnich1,Gonzalez,Costa,Fareghbal2,Krishnan}\footnote{A free field realization of this algebra was first obtained
in \cite{Banerjee1}}. Similar results have been obtained for supersymmetric theories in asymptotic AdS$_3$ spaces \cite{Coussaert,Banados,Justin}.  \\
The enhancement of the symmetry algebra of flat three dimensional gravity has 
been extended to the $N=1$ \cite{Barnich2} and  $N=2$ \cite{Lodato} supersymmetric 
cases. In \cite{BLN1} all possible $N-$extended quantum Super BMS$_3$ algebras were 
found as a well-defined In\"on\"u-Wigner  contraction of the super-Virasoro  algebras. 
The $N=4$ and  $N=8$ algebras also possess non trivial U$(1)$ and non-abelian 
internal $R-$symmetries. The scaling proposal for the R-charges was the main ingredient of this construction.\\
The purpose of the present paper is to find the $N=4$ Super BMS$_3$ algebra i.e. the algebra of three dimensional $N=4$ flat supergravity theory at null infinity, by a direct asymptotic symmetry analysis \emph{a' la} \cite{Barnich2}, i.e. by finding the appropriate boundary conditions to impose on the fields. This provides a check of the algebra found in \cite{BLN1} while also validating the proposed scaling of the $R-$charges.  We leave the similar analysis for the $N=8$ Super BMS$_3$ to a future project \cite{BLN3}.\\
The $N=4$ Super BMS$_3$ algebra obtained, which is the  central result of  this paper, is given in equation 
\ref{flatalgebra}. The agreement with the result of  \cite{BLN1}, as will be clear from the detailed analysis done in later sections, works out it a non trivial way. In fact, it was noticed  long back in \cite{HenneauxMaoz} that the presence of R-symmetry in the extended superconformal algebra leads to non-linearities in the asymptotic symmetry algebra. Those non-linearities can be canceled only by appropriate Sugawara shift of the stress tensor. We will explicitly show how this issue arises when computing the asymptotic AdS 
algebras and its In\"on\"u-Wigner contraction which gives the $N=4$ Super BMS$_3$ algebra \ref{flatalgebra}.

The paper is organized as follows: in the second section, we present the action for three 
dimensional $N=4$ supergravity theory. In the next section, we derive the $N=4$ Super 
BMS$_3$ algebra by choosing the appropriate boundary conditions  for the fields.
In section 4, we present the bounds on the energy of asymptotically flat solutions of the theory, imposed by supersymmetry. We also solve the asymptotic and global Killing spinor equations, and provide explicit solutions. Finally in the last section, we show how the asymptotic algebra is obtained by an appropriate flat limit of the asymptotic AdS$_3$ algebra. 
Our notations, conventions and some details of the computations are presented in the appendices.

\section{Construction of the action}
In three space-time dimensions, a gravity theory with (non-)zero cosmological constant possesses a Chern-Simons formulation. For a three dimensional gauge field $\cA= \cA_{\mu}dx^{\mu}$, the Chern-Simons action is given by,
\begin{equation}\label{csaction}
I [\cA] = \frac{k}{4 \pi}\int \langle \cA, d\cA+ \frac{2}{3}\cA^2 \rangle,
\end{equation}
where $\langle, \rangle $ denotes the invariant bilinear form that one constructs from the symmetry algebra of the corresponding theory (see Appendix \ref{appc} for details on how to build this bilinear form). \\

As mentioned in the introduction, in this paper we want to construct the asymptotic symmetry algebra of three dimensional $N=4$ flat supergravity theory. 
The bulk symmetry algebra for this theory consists of bosonic generators $\cJ_a, P_a , (a=0,1,2), \cR$, $\cS$ and  Majorana fermionic generators 
$\cQ^{1\,\pm}_{\alpha}, \cQ^{2\,\pm}_{\alpha}, (\alpha= \pm \frac{1}{2})$. The commutation relations are:
\begin{align}
[\cJ_a,\cJ_b] & = \epsilon_{abc}\cJ^c\,,\qquad [\cJ_a, P_b]  = \epsilon_{abc}P^c\,,\qquad  [\cJ_a,\cQ^{1,2\,\pm}_{\alpha}]& = \frac12(\Gamma_a)^{\beta}{}_{\alpha} \cQ^{1,2\,\pm}_{\beta}\,, 
\nonumber\\
[\cR,\cQ^{1\,\pm}_\a]&=\pm\tfrac12\, \cQ^{1\,\pm}_\a \,, \qquad  [\cR,\cQ^{2\,\pm}_\a]=\mp\tfrac12\, \cQ^{2\,\pm}_\a 
\nonumber\\
\{\cQ^{1\pm}_{\alpha}, \cQ^{1\mp}_{\beta} \} & = - \frac12 (C\Gamma^a)_{\alpha\beta}P_a\mp \tfrac12\,C_{\a\b}\,\cS,, 
\nonumber\\
 \{\cQ^{2\pm}_{\alpha}, \cQ^{2\mp}_{\beta} \} & = - \frac12 (C\Gamma^a)_{\alpha\beta}P_a \pm \tfrac12\,C_{\a\b}\,\cS\,.\label{QQPdem}
\end{align}
Here, $\cS$ is a possible central extension of the super Poincare algebra while $\cR$ acts as a proper R-symmetry.  One can construct the invariant non-degenerate bilinear form for this algebra (see Appendix \ref{appc}) whose  non-zero elements are \footnote{Our conventions are summarized in appendix \ref{appa}.} ,

\begin{equation}
\langle \cJ_a , P_b \rangle=\eta_{ab}\;,\qquad \langle \cQ^{1,2\pm}_a , \cQ^{1,2 \mp}_\b\rangle=C_{\a\b}\;,\qquad \langle \cR , \cS \rangle=-1
\end{equation}

To write down the action for this supergravity theory, one expands the gauge field in terms of the basis generators as ,
\begin{equation}\label{flatA}
\cA=e^aP_a+\o^a\,\cJ_a+\sum_{\a=\pm}\psi^{1\a}_\pm \cQ^{1\pm}_\a+\sum_{\a=\pm}\psi^{2\a}_\pm \cQ^{2\pm}_\a+\upsilon \cR+\sigma\cS,
\end{equation}
where, $e^a$ is the vielbein field, $\o^a$ is the corresponding dual spin connection, $\psi^{1\a}_\pm,\psi^{2\a}_\pm$ are Majorana gravitini and $\upsilon, \sigma$ 
are internal gauge fields.
With this, we can readily write down the action for $N=4$ asymptotically flat Supergravity theory as,
\begin{align}
\label{eq:flat_action}
S&=\frac{k}{4\pi}\int 2 e^a\,R_a-\sigma{\rm d}\upsilon-\upsilon {\rm d}\sigma+\sum_{a=\pm} \bar\psi^1_a D\psi^1_{-a}+\sum_{a=\pm} \bar\psi^2_a D\psi^2_{-a}
\end{align}
where 
\begin{align}
D\psi^{1}_\pm&={\rm d}\psi^{1}_\pm+\tfrac12\o^a\,\G_a\psi^{1}_\pm\pm\tfrac12\upsilon\psi^{1}_\pm \;,
\nonumber\\
D\psi^{2}_\pm&={\rm d}\psi^{2}_\pm+\tfrac12\o^a\,\G_a\psi^{2}_\pm\mp\tfrac12\upsilon\psi^{2}_\pm \;,
\\
R^a&={\rm d}\o^a+\tfrac12\,\varepsilon^a{}_{bc}\,\o^b\,\o^c \;,
\end{align}
The invariance of the  action  $S$ under the supersymmetry \ref{QQPdem} can be straightforwardly checked by using the transformations, 
$$\delta\cA={\rm d}\l^{susy}+[\cA,\l^{susy}],
 \quad \l^{susy}=\theta_\pm^{1\a}\cQ^{1\pm}_\a+\theta^{2\a}_{\pm}\cQ^{2\pm}_\a.$$
which explicitly read:
\begin{align*}
\delta e^a&=\tfrac12 (\bar\theta^1_+\G^a\psi^1_-+\bar\theta^1_-\G^a\psi^1_++\bar\theta^2_+\G^a\psi^2_-+\bar\theta^2_-\G^a\psi^2_+)\;, \qquad \delta\o^a =0\;,
\nonumber\\
\delta\psi^{1\a}_\pm&={\rm d}\theta^{1\a}_\pm+\tfrac12\,\o^a\,\G_a\,\theta^{1\a}_\pm\pm\tfrac12\upsilon\theta^{1\a}_\pm=D\theta^{1\a}_\pm\;, 
\nonumber\\
\delta\psi^{2\a}_\pm&={\rm d}\theta^{2\a}_\pm+\tfrac12\,\o^a\,\G_a\,\theta^{2\a}_\pm\mp\tfrac12\upsilon\theta^{2\a}_\pm=
D\theta^{2\a}_\pm \;, 
\nonumber\\
\delta\sigma&=\mp\tfrac12(\bar\psi^1_\pm\theta^1_\mp-\bar\psi^2_\pm\theta^2_\mp)\;,\qquad\delta\upsilon=0\;.
\end{align*}
The algebra of supersymmetry closes on-shell into a general coordinate transformation, a Lorentz transformation (with dualized parameter $\lambda^a = \epsilon^{abc}\Lambda_{bc}$) and a supersymmetry transformation with parameters $\varepsilon_\pm=-\xi^\nu\,\psi^1_{\nu\,\pm}$ and $\vartheta_\pm=-\xi^\nu\,\psi^2_{\nu\,\pm}$: 
\begin{align}
& [\delta(\varepsilon^1_+,\varepsilon^1_-,\vartheta^1_+,\vartheta^1_-),\delta(\varepsilon^2_+,\varepsilon^2_-,\vartheta^2_+,\vartheta^2_-)]= 
\delta_{\rm  Lor} (\lambda^a = - \xi^{\nu}\omega_{\nu}{}^a) + \delta_{\rm susy}(\varepsilon_+,\varepsilon_-, \vartheta_+,\vartheta_-) \nonumber \\
 & \qquad +\delta_{\rm g.c.}(\xi^{\nu} = -\tfrac12 (\bar\varepsilon^2_- \Gamma^{\nu}\varepsilon^1_+ + \bar\varepsilon^2_+ \Gamma^{\nu}\varepsilon^1_-+\bar\vartheta^2_- \Gamma^{\nu}\vartheta^1_+ + \bar\vartheta^2_+ \Gamma^{\nu}\vartheta^1_-)).
\end{align}
The dynamical equations  are:
\begin{align*}
T^a&=-\tfrac12\,(\bar\psi^1_+\,\G^a\psi^1_-+\bar\psi^2_+\,\G^a\psi^2_-)\;\quad D\psi^{1,2}_\pm=D\bar\psi^{1,2}_\pm=0\;,
\nonumber\\
{\rm d}\upsilon&=F_{\upsilon}=0 \qquad \qquad 2\,F_{\sigma} +(\bar\psi^1_-\,\psi^1_+ -\bar\psi^2_+\,\psi^2_-)=0\;,
\end{align*}  
with the  torsion tensor $T^a={\rm d}e^a+\varepsilon^a{}_{bc}\o^b\,\o^c$. \\
In this paper, we are interested in finding the asymptotic symmetry algebra for the above theory at null infinity. To do so, we change frame for the generators, as  was done in \cite{Lodato}. The new generators $\{M_n,\cL_n,q^{1,2\pm}_\a,\cR,\cS\}$ are related to the previous ones by the following relations \footnote{Our conventions are summarized in appendix \ref{appa} and appendix \ref{appd}.}:
\begin{align}
M_n&=P_a U^a_n\;,\quad \cL_n=\cJ_a U^a_n\;,\quad q^{1\pm}_\a=\sqrt{2}\cQ^{1\pm}_\a\;,\quad q^{2\pm}_\a=\sqrt{2}\cQ^{2\pm}_{\a}\;,\nonumber
\end{align}
with $(\cR, \cS)$ remaining unchanged. In terms of these generators, the super Poincare  algebra reads as:
\begin{align}
\label{eq:Vir_alg_new}
[\cL_n,\cL_m]&=(n-m)\cL_{n+m}\;,\qquad [\cL_n,M_m]=(n-m)M_{n+m}\;,\qquad [M_n,M_m]=0\;,
\nonumber\\
[\cL_n,q^{1,2\,\pm}_\a]&=\big(\frac{n}{2}-\a\big)\,q^{1,2\,\pm}_{n+\a}\;,\qquad [M_n,q^{1,2\pm}_\a]=0\;,\qquad  [\cS,q^{1,2\pm}_\a]=0\;,
\nonumber\\
[\cR,q^{1\pm}_\a]&=\pm\tfrac12\,q^{1\pm}_\a\;,\qquad\qquad\quad
[\cR,q^{2\pm}_\a]=\mp\tfrac12\,q^{2\pm}_\a\;,
\nonumber\\
\{q^{1\pm}_{\alpha},q^{1\mp}_{\beta}\}&=M_{\alpha+\beta}\pm(\a-\b)\cS,\qquad \{q^{2\pm}_{\alpha},q^{2\mp}_{\beta}\}=M_{\alpha+\beta}\mp(\a-\b)\cS\;.
\end{align}
In the next section, we find the right asymptotic gauge field to finally arrive at the asymptotic symmetry group for this three dimensional $N=4$ flat super-gravity.

\section{$N=4$  BMS$_3$ asymptotic algebra}

The aim of this paper is to find the asymptotic symmetry algebra for a specific set of  boundary conditions of  the gauge field. The procedure is well defined and has been used in the literature, for both asymptotically flat and $AdS$ theories. 
The boundary conditions need to (i) extend the ones of the purely gravitational sector so as to include the bosonic solutions of interest, mentioned in the previous section, and (ii) relaxed enough so as to enlarge the set of asymptotic symmetries from $N=4$ super-Poincare to its $N=4$ super-BMS extension. Obviously, they also fix the form of the metric which is, in the usual BMS gauge with Eddington-Finkelstein coordinates $(u,r,\varphi)$:
\begin{equation}
{\rm d}s^2=\eta_{ab}\,e^a\,e^b=\cM\,{\rm d}u^2-2{\rm d}u\,{\rm d}r+\cN\,{\rm d}u\,{\rm d}\varphi+r^2\,{\rm d}\varphi^2
\end{equation}
The gauge fields at the boundary is hence chosen in a radial gauge,
\begin{equation}
\cA= b^{-1}(a+d)b\;,\qquad b=\exp\big(\frac{r}{2}M_{-1}\big)
\end{equation}
where now $a(u,\varphi)=a_\varphi{\rm d}\phi+a_u{\rm d}u$ reads:
\begin{align}
\label{eq:gauge_field_flat}
a_u
=&M_1-\tfrac14\,\cM\,M_{-1}-\frac{i \rho}{2}\,\cS
\nonumber\\
a_\varphi=&\cL_1-\tfrac14\,\cM\,\cL_{-1}-\tfrac{1}{4}\cN\,M_{-1}-\frac{i\phi}{2}\,\cS-
\frac{i \rho}{2}\,\cR
\nonumber\\
&-\frac{1}{4}\big(\Psi^1_+\,q^{1 +}_- -\Psi^1_-\,q^{1-}_-\big)+\frac{1}{4}\big(\Psi^2_+\,q^{2+}_--\Psi^2_-\,q^{2 -}_-\big)
\end{align}
The various charges appearing in the above expression asymptotically will 
only have $u$ and $\varphi$ dependence. 
The asymptotic symmetries correspond to the set of gauge transformations that preserve this behaviour together with the dynamical equations:
\begin{equation}
{\rm d}a+ \frac{1}{2 }[a,a]=0, \qquad \delta a = {\rm d}\L+[a,\L],
\end{equation}
where, the parameter $\Lambda$ is Lie-algebra valued and depends on various arbitrary functions of $u$ and $\varphi$,
 \begin{equation}
\L=\U^n\,\cL_n+\xi^n\,M_n+\z^{1\,\a}_{+}\,q^{1\,+}_\a+\z^{1\,\a}_{-}\,q^{1\,-}_\a+\z^{2\,\a}_{+}\,q^{2\,+}_\a+\z^{2\,\a}_{-}\,q^{2\,-}_\a+\l_\cR\,\cR+\l_\cS\,\cS.
\end{equation}
From the equations of motion one gets the following differential identities:
\begin{align}
\label{eq:ident_1}
\partial_\varphi\cM&=\partial_u\cN\;, \quad  \partial_u\cM=0\;\quad \partial_u\,\rho=0\;,
\\
\label{eq:ident_2}
\partial_\varphi \rho&=\partial_u\phi\;,\quad \partial_u\Psi^1_\pm=0\;,\quad \partial_u\Psi^2_\pm=0\;.
\end{align}
Similar identities exist also for the parameters, 
\begin{align}
\label{eq:ident_3}
&\partial_u\xi^n=\partial_\varphi\U^n\;,\quad \partial_u\U^n=0\;,\quad \partial_u\z^{1\a}_\pm=0\;,\quad 
\\
\label{eq:ident_4}
& \partial_u\z^{2\a}_\pm=0\;,\qquad\partial_u\l_\cS=\partial_\varphi\l_\cR\;,\quad \partial_u\,\l_\cR=0\;.
\end{align}
Here, we see that fields and parameters are not independent of each other.\\
Next we start the analysis of the gauge variation condition,  which constrain the parameters  even further:  
\begin{align}
\label{eq:ident_5}
\xi^0&=-\partial_\varphi\xi^++r\,\U^+,\qquad \U^0=-\partial_\varphi\U^+\;, \quad \U^-=\tfrac12\partial_\varphi^2\U^+-\tfrac14\,\cM\,\U^+  =0 
\\
\nonumber\\
\label{eq:ident_6}
\xi^-=&\tfrac12\partial_\varphi^2\xi^+ -\tfrac14\,\big(\cM\,\xi^++\cN\U^+\big)-\tfrac18\big(\Psi^1_+\z^{1+}_--\Psi^1_-\z^{1+}_+\big) +\tfrac18\big(\Psi^2_+\z^{2+}_--\Psi^2_-\z^{2+}_+\big)
\end{align}
where we made multiple use of the above identities. 
The constraints on the fermionic parameters are: 
\begin{align*}
\z^{1-}_\pm&=-\partial_\varphi\z^{1+}_\pm \mp\frac{1}{4}\Psi^1_\pm\U^+\pm
\frac{i}{4}\,\rho\,\z^{1+}_\pm  \nonumber \\
\z^{2-}_\pm &=-\partial_\varphi\z^{2+}_\pm\pm\frac{1}{4}\Psi^2_\pm\U^+\mp\frac{i}{4}\,\rho\,\z^{2+}_\pm
\end{align*}
From now on, then we will use $\zeta^{1,2 +}_\pm=\zeta^{1,2}_\pm$. 
Finally we write down the variation of the fields. For bosonic fields, we get,
\begin{align}
\delta\cM&=-2\partial_\varphi^3\U^++2\cM\partial_\varphi\U^++\partial_\varphi\cM\U^+\;,
\nonumber\\
\delta\cN&=-2\partial_\varphi^3\xi^++2\cM\partial_\varphi\xi^+
+2\cN\partial_\varphi\U^++\partial_\varphi\cM\xi^+\;,
+\partial_\varphi\cN\U^+ 
\nonumber\\
&\quad+\tfrac12\big(\partial_\varphi\Psi^1_+\z^{1}_-+3\,\Psi^1_+\partial_\varphi\z^{1}_-
-\partial_\varphi\Psi^1_-\z^{1}_+-3\,\Psi^1_-\partial_\varphi\z^{1}_+ 
+\frac{i}{2}\Psi^1_+\z^{1}_-\rho+\frac{i}{2}\Psi^1_-\z^{1}_+\rho\big)
\nonumber\\
&\quad-\tfrac12\big(\partial_\varphi\Psi^2_+\z^{2}_-+3\,\Psi^2_+\partial_\varphi\z^{2}_-
-\partial_\varphi\Psi^2_-\z^{2}_+-3\,\Psi^2_-\partial_\varphi\z^{2}_+ 
-\frac{i}{2}\Psi^2_+\z^{2}_-\rho-\frac{i}{2}\Psi^2_-\z^{2}_+\rho\big)\;, 
\nonumber \\
\delta\phi&=2i\partial_\varphi\l_\cS-\frac{\rm  i}{2}(\Psi^1_+\,\z^{1}_-+\Psi^1_-\,\z^{1}_+)
-\frac{\rm i}{2}(\Psi^2_+\,\z^{2}_-+\Psi^2_-\,\z^{2}_+)\;,
\nonumber\\
\delta\rho&=2i\partial_\varphi\l_\cR, \quad 
\end{align}
For the fermionic fields we get:
\begin{align}
\label{eq:rgav_variat}
\delta\Psi^1_\pm&=\pm4\partial_\varphi^2\z^{1 }_\pm +\Big(\partial_\varphi\Psi^1_\pm \U^++\tfrac32\Psi^1_\pm\partial_\varphi\U^+\Big)
- {\rm i}\Big(\partial_\varphi\,\rho\z^{1}_\pm+2\rho\partial_\varphi\z^{1}_\pm\Big)\mp\cM\z^{1}_\pm
\nonumber\\
&\qquad \mp\frac{{\rm i}}{4}\Psi^1_\pm\rho\U^+\mp\frac{1}{2}\l_\cR\Psi^1_\pm\mp\frac{1}{4}\rho^2\z^{1}_\pm\;, 
\nonumber \\
\delta\Psi^2_\pm&=\mp4\partial_\varphi^2\z^{2 }_\pm +\Big(\partial_\varphi\Psi^2_\pm \U^++\tfrac32\Psi^2_\pm\partial_\varphi\U^+\Big)
- {\rm i}\Big(\partial_\varphi\rho\,\z^{2}_\pm+2\rho\,\partial_\varphi\z^{2}_\pm\Big)\pm\cM\z^{2}_\pm
\nonumber\\
&\qquad \pm\frac{ {\rm i}}{4}\Psi^2_\pm\rho\U^+\pm\frac{1}{2}\l_\cR\Psi^2_\pm\pm\frac{1}{4}\rho^2\z^{2}_\pm.
\end{align}
￼￼The variation of the canonical generators that corresponds to the asymptotic symmetries of this theory spanned by fields can be obtained in the canonical approach . In the case of a Chern-Simons theory in three dimensions, they are given by, 
\begin{equation}
\delta\cC =-\frac{k }{2\pi}\int \langle \L, \delta\cA_\varphi\rangle d\varphi
\end{equation}
This expression is linear in the fields variations and it reads explicitly
\begin{align}\label{crgvariation}
\delta\cC&=-\frac{k}{4\pi}\int \big(\U^+\delta\mathfrak{J}+T\d\cM+\delta\Psi^1_+ \z^{1}_-
-\delta\Psi_-^1\z^{1}_+
\nonumber\\
&\qquad\qquad-\delta\Psi^2_+ \z^{2}_-+\delta\Psi_-^2\z^{2}_+
+i\l_\cR\delta\phi+i\l_\cS\delta\rho\big) {\rm d}\varphi\;,
\end{align}
where we have used the supertraces suitable for the current basis (derived from relations in appendix \ref{appc} and appendix \ref{appd}):
\begin{equation}
\langle \cL_n, M_m \rangle = \g_{nm} \;,\qquad\langle q^{1,2\pm}_\a,q^{1,2 \mp}_\b \rangle=2\,C_{\a\b}\;,\qquad \langle \cR,\cS \rangle=-1\;,
\end{equation} 
and solved \eqref{eq:ident_1} and \eqref{eq:ident_3}:
\begin{equation}
\cN=\mathfrak{J}(\varphi)+u\partial_\varphi\cM, \quad  \xi^+=T(\varphi)+u\partial_\varphi \U^+.
\end{equation}
Under some mild regularity assumptions for  the variations, we can readily read off the charge from the above variation formula as,
\begin{align}
\cC&=-\frac{k}{4\pi}\int \big(\U^+\mathfrak{J}+T\cM+\Psi^1_+ \z^{1}_-
-\Psi_-^1\z^{1}_+
\nonumber\\
&\qquad\qquad-\Psi^2_+ \z^{2}_-+\Psi_-^2\z^{2}_+
+i\l_\cR\phi+i\l_\cS\rho\big) {\rm d}\varphi
\nonumber\\
&=-\frac{2}{k}\sum_n \U^+_{-n}\mathfrak{J}_n+T_{-n}\cM_n+\Psi^{1+}_n \z^{1 -}_{-n}-\Psi_n^{1-}\z^{1 +}_{-n}-\Psi^{2+}_{n} \z^{2-}_{-n}+\Psi_n^2\z^{2 +}_{-n}
\nonumber\\
&\qquad\qquad+{\rm i}\l^\cR_{-n}\cR_n+{\rm i}\l^\cS_{-n}\cS_n
\end{align}
where $\z^{1,2 \pm}_n$ are now the modes of $\z^{1,2}_\pm$.
We can derive the asymptotic symmetry algebra of this configuration using the asymptotic charge and its variation as given above. In
particular,  the Poisson brackets among various modes of the 
fields can be obtained using the formula  
\begin{equation}
\{\cC[\l_1],\cC[\l_2]\}_{PB}=\delta_{\l_1}\cC[\l_2]\;.
\end{equation} 
It reads:
\begin{align}
\label{eq:BMS_1_PB}
{\rm i}\{\mathfrak{J}_n,\mathfrak{J}_m\}&=(n-m)\mathfrak{J}_{n+m},\quad 
{\rm i}\{\mathfrak{J}_n,\cM_m\}=(n-m)\cM_{n+m}+\frac{c_M}{12}\,n^3\delta_{n+m,0}
\nonumber\\
{\rm i}\{\mathfrak{J}_n,\Psi^{1,2\pm}_r\}&=
\big(\frac{n}{2}-r\big)\Psi^{1,2\pm}_{r+n}\mp \tfrac14 [\Psi^ {1,2\pm} \cS]_{n+r} 
\nonumber\\
{\rm i}\{\cR_n,\Psi^{1\pm}_r\}&=\pm\tfrac12 \Psi^{1\pm}_{r+n},\quad 
{\rm i}\{\cR_n,\Psi^{2\pm}_r\}=\mp\tfrac12 \Psi^{2\pm}_{r+n},\quad 
{\rm i}\{\cR_n,\cS_m\}=\frac{c_M}{12}\, n \delta_{n+m,0}
\nonumber\\
\{\Psi^{1+}_r,\Psi^{1-}_s\}&=\cM_{r+s}+(r-s)\cS_{r+s}+\tfrac14 [\cS\cS]_{r+s}+\frac{c_M}{6}\,r^2\delta_{r+s,0}
\nonumber\\
\{\Psi^{2+}_r,\Psi^{2-}_s\}&=\cM_{r+s}-(r-s)\cS_{r+s}+\tfrac14 [\cS\cS]_{r+s}+\frac{c_M}{6}\,r^2\delta_{r+s,0},
\end{align}
where $c_M=12k$ and the modes are defined as follows:
\begin{align}\label{bmsmodes}
\mathfrak{J}_n&=\frac{k}{4\pi}\int d\varphi e^{in\varphi}\mathfrak{J} , \qquad 
\mathcal{M}_n=\frac{k}{4\pi}\int d\varphi e^{in\varphi}\mathcal{M},  \nonumber \\
\mathcal{R}_n&=\frac{k}{4\pi}\int d\varphi e^{in\varphi}\phi,  \qquad
\mathcal{S}_n=\frac{k}{4\pi}\int d\varphi e^{in\varphi}\rho,  \nonumber \\
\Psi^{1,2\pm}_r&=\frac{k}{4\pi}\int d\varphi e^{ir\varphi}\Psi^{1,2\pm},  \qquad
[\Psi^{1,2\pm}\mathcal{S}]_r=\frac{k}{4\pi}\int d\varphi e^{ir\varphi}\Psi^{1,2\pm}\rho,  \nonumber \\
[\mathcal{S}\mathcal{S}]_{\alpha}&=\frac{k}{4\pi}\int d\varphi e^{i\alpha \varphi}\rho\rho,  \qquad
\delta_{n,0}=\frac{1}{2\pi}\int d\varphi e^{in\varphi}. 
\end{align}
and similarly for the parameters. To get the Poisson brackets, we need the inverse relations among the fields and the modes as well. 
For example, for fields $\mathfrak{J}(\varphi)$ and 
$\mathcal{M}(\varphi)$, the inverse relations are given by:
\begin{equation}
\mathfrak{J}(\varphi)=\frac{2}{k}\sum_{n}e^{-in\varphi}\mathfrak{J}_n,  \qquad \mathcal{M}(\varphi)=\frac{2}{k}\sum_{n}e^{-in\varphi}\mathcal{M}_n .
\end{equation}
Here we notice that the Poisson bracket $\{\mathfrak{J}_n,\psi^{1,2,\pm}_r\}$ contains a spurious 
term (the last term) while $i\{\mathfrak{J}_n,\mathcal{R}_m\}$ is zero. Also, the 
Poisson bracket $\{\Psi^{1,2+},\Psi^{1,2-}\}$ contains a quadratic term in the $\cS$ generator. Hence, at 
this stage, the algebra looks quite different from the one derived in \cite{BLN1}. The resolution of these differences resides on a simple argument: since we are dealing with a theory with one internal U$(1)$ symmetry, the physical energy-momentum tensor should have a contribution from the corresponding U$(1)$ current. Thus it 
is important to add a 
Sugawara-like term to $\mathfrak{J}_n$  as follows:
\begin{equation}
\hat{\mathfrak{J}}_n=\mathfrak{J}_n+\tfrac12(\mathcal{R}\mathcal{S})_n\;;
\end{equation}  
With these shifts of the modes, some spurious terms get canceled or absorbed and the new Poisson brackets read :
\begin{align}
i\{\hat{\mathfrak{J}}_n,\psi^{1,2\pm}_r\}&=\left(\frac{n}{2}-r\right)\psi^{1,2\pm}_{r+n}\;, 
\nonumber\\ 
i\{\hat{\mathfrak{J}}_n,\mathcal{R}_m\}&=-m\mathcal{R}_{m+n}
\nonumber \\
i\{\hat{\mathfrak{J}}_n,\mathcal{S}_m\}&=-m\mathcal{S}_{m+n}
\end{align}
Finally, we also perform a shift on $\cM_n$ 
\begin{equation}
\hat{\cM}_n=\cM_n+\tfrac14 [\cS\cS]_n 
\end{equation}
to absorb the quadratic term in $\cS$ in the $\{\Psi,\Psi\}$ bracket. Also:  
 $$i \{\hat{\cM}_n,\cR_m\}=-m\,\cS_{n+m}.$$\\
 The underlying justification for these shifts will become clear in the last section.
In the next subsection, instead we present the final result for  the $N=4$ super-BMS algebra which is in agreement with \cite{BLN1}.
 
 \subsection{The BMS Algebra}
 \label{sec:BMSalgebra}
 Here, we present the main result of this paper, the algebra of three
 dimensional $N=4$ flat super-gravity at null infinity, 
namely the $N=4$ Super BMS$_3$ algebra.
 From the non-zero Poisson brackets, we can write down the final form of the algebra. 
 The rule that we follow is : 
 $$i \{,\}_{PB}\rightarrow [,] \,\,\,\,\,\, \text{and} \quad \{,\}_{PB}\rightarrow 
 \{,\}\;.$$
The algebra in terms of 
commutators and anti-commutators is given by:
\begin{align}
\label{flatalgebra}
[\hat{\mathfrak{J}}_n,\hat{\mathfrak{J}}_m\}&=(n-m)\hat{\mathfrak{J}}_{n+m}
+\frac{c_J}{12}\,n^3\delta_{n+m,0},\quad
[\cR_n,\cR_m]=\frac{c_J}{12}\, n\, \delta_{n+m,0} 
\nonumber \\  
[\hat{\mathfrak{J}}_n,\hat{\cM}_m\}&=(n-m)\hat{\cM}_{n+m}+\frac{c_M}{12}\,n^3\delta_{n+m,0},\quad
[\cR_n,\cS_m]=\frac{c_M}{12}\, n\, \delta_{n+m,0}
\nonumber\\
[\hat{\mathfrak{J}}_n,\cR_m]&=-m\,\cR_{n+m}\;,\quad[\hat{\mathfrak{J}}_n,\cS_m]=-m\,\cS_{n+m}\;,
\quad[\hat{\cM}_n,\cR_m]=-m\,\cS_{n+m}
\nonumber\\
[\hat{\mathfrak{J}}_n,\Psi^{1,2\pm}_r\}&=\big(\frac{n}{2}-r\big)\Psi^{1,2\pm}_{r+n},\quad
[\cR_n,\Psi^{1\pm}_r]=\pm\tfrac12 \Psi^{1\pm}_{r+n}, \quad 
[\cR_n,\Psi^{2\pm}_r]=\mp\tfrac12 \Psi^{2\pm}_{r+n}, 
\nonumber\\
\{\Psi^{1+}_r,\Psi^{1-}_s\}&=\hat{\cM}_{r+s}+(r-s)\cS_{r+s}+\frac{c_M}{6}\,r^2\delta_{r+s,0}
\nonumber\\
\{\Psi^{2+}_r,\Psi^{2-}_s\}&=\hat{\cM}_{r+s}-(r-s)\cS_{r+s}+\frac{c_M}{6}\,r^2\delta_{r+s,0}
\end{align}
Here, we have presented the most generic possible quantum extension of the algebra by 
allowing a possible central extension to the $[\hat{\mathfrak{J}}_n,\hat{\mathfrak{J}}_m]$ 
and $[\mathcal{R}_n,\mathcal{R}_m]$ commutator.  We 
also notice that, after adding suitable shifts to $\mathfrak{J}_n$ and $\cM_n$, we obtain the same algebra presented in \cite{BLN1}.

\section{Energy bound and  Killing spinors}
In this section, we look for the energy bounds for three dimensional
$N=4$ asymptotically flat supergravity theories.  We find the
asymptotic symmetries that leave the asymptotic background
unchanged. Finally, we find the global killing spinors for this system.

\subsection{Supersymmetry Energy  bound}
As it is well-know, supersymmetry imposes constraints on the energy of
supersymmetric states.  One can find it from the super algebra.
Specifically for our case, considering anti-periodic boundary
conditions on the fermions\footnote{We have not studied the Ramond
  boundary conditions for the fermions, more can be found
in \cite{Witten,Horowitz,Barnich2}. }, we see that the global part of
the algebra consists of 
$(\mathfrak{J}_m, \hat{\cM}_m,\Psi^{1,2\pm}_r, \cR)$, where $m=-1,0,1$
and $r=\pm \frac12$.  For the quantum theory, following \cite{DeserT, DeserA,CoussaertH}, we consider all possible positive-definite 
combinations of  the supercharges $\Psi^{1,2 \pm}_{\pm1/2}$ and get:
\begin{equation}
\hat{\cM}_0=\tfrac14\sum_{\substack{i=1,2 \\ \a=\pm1/2}}\Psi^{i +}_\a\,\Psi^{i -}_{-\a}+\Psi^{i -}_{-\a}\,\Psi^{i +}_\a-\frac{k}{2}\ge -\frac{k}{2}=-\frac{1}{8G}
\end{equation}
Here, it is important to note that, we have derived the above bound
for $\hat{\cM}_0$, the shifted charge. 
Unlike ${\cM}_0$, the latter  
satisfies this nicer bound. This implies that, for extended supersymmetric
cases, the right physical charge at null infinity corresponds to $\hat{\cM}$ 
It is also very clear that for the Minkowski vacuum for which $\hat{\cM}_0=\cM_0=-\frac{1}{8G}$ as all the other fields, including the R- and S-symmetry gauge  fields are vanishing, the bound is saturated. Hence, Minkowski space is certainly a ground state for this theory.
\subsection{Asymptotic Killing Spinors}

To study the asymptotic supersymmetries that preserve the
asymptotically flat backgrounds, we impose that both the gravitinos
and their generic variation be zero, at infinity. This is known as ``asymptotic Killing spinor equation". One hence has  to solve the 
simplified version of equations \eqref{eq:rgav_variat}, i.e.:
\begin{equation}
\partial_\varphi^2\zeta^{i}_\pm \mp\frac{\rm i}{2}\rho\,\partial_\varphi\zeta^{i}_\pm-\tfrac14(\cM+\tfrac14\,\rho^2)\zeta^i=0
\end{equation}
where $i=1,2$ and we assumed $\partial_\varphi \rho=0$ and $\cM$ constant. The general solutions to the above equations read:
\begin{align}
\zeta^i_+&= e^{-{\rm i}\tfrac{\rho}4\,\varphi}\big(c^i_1\,e^{\tfrac{\sqrt{\cM}}{2}\varphi}+c^i_2\,e^{-\tfrac{\sqrt{\cM}}{2}\varphi}\big)
\nonumber\\
\zeta^i_-&= e^{{\rm i}\tfrac{\rho}4\,\varphi}
\big(d^i_1\,e^{\tfrac{\sqrt{\cM}}2\,\varphi}+d^i_2\,e^{-\tfrac{\sqrt{\cM}}2\,\varphi}\big)
\end{align}
for arbitrary $c^i_{1,2}$ and $d^i_{1,2}$ constant spinors.
The solutions are well-defined, given the periodicity of $\varphi$
only when $\cM=-n^2$ and $n>0$, a strictly positive integer without
loss of  generality. \\
For $n=1,\rho=0$ we find the Killing spinors for the Minkowski
vacuum, $\cM=-1$. For  $n > 1$, the energy bound is violated and we
have angular defect solutions \cite{Miskovic:2009uz}.

 \subsection{Global Killing vectors}
We end this section with the study of global killing spinors. These
describe globally defined supersymmetry transformations that leave
the pure bosonic solution in the asymptotic region
invariant. Depending on the range of the mass parameter, the pure
bosonic zero mode solutions include cosmological solutions \cite{Cornalba1,Cornalba2}, stationary
conical defects solutions \cite{Miskovic:2009uz}, the
Minkowski spacetime and angular excess solutions of \cite{DeserJ,BarnichGG}. The global Killing spinor equations is
given as,
\begin{equation}
\label{eq:glob_kill_spin}
D \zeta^{1}_\pm=({\rm d}+\o\pm\tfrac12\upsilon)\zeta^{1}_\pm =0
\end{equation}
From the gauge field \eqref{eq:gauge_field_flat}, we obtain the values of  the spin connection and the R-gauge field:
\begin{equation}
\o=\tfrac12\o^n\,\tilde\G_n=\L^{-1}{\rm d}\L  \;,\qquad \Lambda =
\exp\left(\frac12 \left(\tilde{\Gamma}_{+1} - 
\frac{\cM}{4}\tilde{\Gamma}_{-1} \right)\varphi \right)\,,
\nonumber\\
\end{equation}
The  general solution of this
equation is obtained from the solution of the homogeneous
 equation ($\upsilon=0$) that was already solved in \cite{Barnich2},
 given as:
\begin{equation}
\zeta^{1}_{\rm hom}=\Lambda^{-1}\zeta^{1}_{0}=\left(\begin{array}{cc}
\cosh\left(\frac{\sqrt{\cM}}{2}\varphi \right) & -\frac{\sqrt{\cM}}{2}\sinh\left(\frac{\sqrt{\cM}}{2}\varphi \right) \\
-\frac{2}{\sqrt{\cM}}\sinh\left(\frac{\sqrt{\cM}}{2}\varphi \right) & \cosh\left(\frac{\sqrt{\cM}}{2}\varphi \right)
\end{array}\right)\zeta^{1}_{0}
\end{equation}
with $\zeta^{1}_{0}$ constant spinors and we  have suppressed the  indices $\pm$.
indices $\pm$. The solution of the inhomogeneous equation with 
non-zero $\upsilon$ is of the form:
\begin{equation}
\zeta^{1,2}_{\pm {\rm gen}}=\Lambda^{-1}(\zeta^{1,2}_{0}+\zeta^{1,2}_\pm(x))
\end{equation}
By explicitly plugging in the above \eqref{eq:glob_kill_spin} we get:
\begin{equation}
{\rm d}\zeta^{1,2}_\pm(x)=\pm\frac{\rm i}2\,\phi {\rm d}\varphi\, (\zeta^{1,2}_0+\zeta^{1,2}_\pm(x))
\end{equation}
where we identified $\upsilon=-{\rm i}\frac{\phi}{2}$ from the form asymptotic gauge field. This differential equation is immediately solved by:
\begin{align}
\partial_r \zeta^{1,2}_\pm&=\partial_u\zeta^{1,2}_\pm=0
\nonumber\\
\zeta^{1,2}_\pm(\varphi)&=\tilde\zeta^1\,e^{\pm{\rm i}\tfrac12\phi\,\varphi}-\zeta^{1}_{0} 
\end{align}
with $\tilde\zeta^1$ constant spinor. Thus the final solution for the global killing spinors takes the form :
\begin{equation}
\zeta^{1}_{\pm {\rm gen}}=\Lambda^{-1}\,\tilde\zeta^{1}\,e^{\mp{\rm i}\tfrac12\phi\,\varphi}
\end{equation}
For the  second Killing spinor, the equation has simply the signs of the R-symmetry gauge field flipped, which corresponds only to the sign of the exponential flipped. Like the asymptotic case, the
Killing spinors are globally well-defined  when $\cM=-n^2$, with $n$
being positive integer. A more detailed discussions can be found in \cite{Barnich2,Lodato,Oblak}.

\section{Super BMS$_3$ as a flat Limit of asymptotically super-AdS$_3$ Supergravity}
It is well-known that the
flat asymptotic algebra can be obtained by taking an appropriate limit (or
contraction) of two copies of the asymptotic AdS algebras. In
\cite{BLN1}, we adopted this limiting procedure to derive all
possible supersymmetric  extensions 
of the BMS$_3$ algebras by considering the limit of the mixed sectors of the superconformal algebra. When the R-symmetry is present, there exist two possible
combinations for the $R-$ charge generators, the democratic and the despotic scalings. The former was 
excluded because the R-generators did not rotate the supercharges, so  that left us with one well-defined combination of the R-symmetry generators of the two super-Virasoro sectors, which led to a $N=4$ super-BMS$_3$ algebra. In 
this paper, we have re-derived this algebra as given in \ref{flatalgebra} by a direct analysis of  the  gauge field boundary conditions. The result is in complete agreement with the results of \cite{BLN1} after 
considering suitable shifts in two generators, as shown in the last section. The 
reason behind these shifts is discussed below. 

As it turns out, the asymptotic AdS algebra considered in \cite{BLN1}
initially contains non-linear terms in $R-$charge generators. 
The asymptotic symmetry algebra of $N-$ extended AdS 
Supergravity theories 
were first discussed in \cite{HenneauxMaoz} and those are not the usual superconformal algebras. For 
completeness, we shall present again those  results, including the Sugawara shifts of the Virasoro generators, by using the Chern-Simons formulation of AdS gravity. 
\subsection{Asymptotic symmetry algebra for (2,0) and (0,2) AdS supergravity}
There are two inequivalent locally supersymmetric extensions of General Relativity 
with negative cosmological constant in three spacetime dimensions containing an R-symmetry,
known as the (2,0) and (0,2) theories. The bulk symmetry algebras 
for both the theories are presented in appendix \ref{appb}.  Here, we
formulate them  as a Chern-Simons theory with appropriate gauge
group Osp$(2|2,\mathbb{R})$.
The action is a  functional of two independent connections $A_+$ and $A_{-}$:
\begin{equation}
I=I[A_+]+I[A_-],
\end{equation}
where, $I[A]$ is defined earlier in \ref{csaction}. Here, we have
defined $x^\pm= u/l\pm \varphi$, where, $l$ is the identical 
AdS radius in both sectors. Hence, the (2,0) sector asymptotically
only depends on $x^+$ and the (0,2) sector depends on $x^-$. The asymptotic behaviour of the gauge fields can be taken to be
\begin{align}\label{gauge1}
A_+&=\Big(L_1+\frac{r}{l}\,L_0+\frac{r^2}{4\,l^2}\,L_{-1}-\tfrac12\,
\mathfrak L_+\,L_{-1}-\tfrac12\,\psi_+\,\cQ^+_-+\tfrac12\,\psi_-\,\cQ^-_-
- i \phi^A_R\, R\Big){\rm d}x^++\frac{{\rm d}r}{2\,l} L_{-1}
\nonumber\\
\bar A_-&=\Big(\bar L_{-1}-\frac{r}{l}\,\bar L_0+\frac{r^2}{4\,l^2}\,\bar L_1-
\tfrac12\,\mathfrak L_-\,\bar L_1-\tfrac12\,\bar\psi_+\,\bar\cQ^+_++\tfrac12\,\bar\psi_-\,
\bar\cQ^-_+- i\bar\phi^A_R\, \bar R\Big){\rm d}x^-+\frac{{\rm d}r}{2\,l}\bar L_1.  
\end{align}
and again from the dynamical equations we get the trivial constraints:
\begin{equation}
\label{eq:constr_2scsectors}
\partial_- \mathfrak L_+=\partial_-\psi_\pm=\partial_-\phi^A_R=0\;, \quad  \partial_+ \mathfrak L_-=\partial_+\bar\psi_\pm=\partial_+\bar\phi^A_R=0\;.
\end{equation}
The asymptotic symmetries for these systems are generated by the asymptotic gauge transformations $\delta A_{\pm} = {\rm d}\L_{\pm}+[A_{\pm},\L_{\pm}]$ for both gauge fields,
 where the transformation parameters are given as,
 \begin{align}
  \L_+&=\chi^n\,L_n+\epsilon^\a_+\,\cQ^+_\a+\epsilon^\a_-\,\cQ^-_\a+\l^A_R\,R\nonumber\\
\L_{-}&=\bar\chi^n\,L_n+\bar\epsilon^\a_+\,\bar\cQ^+_\a+\bar\epsilon^\a_-\,\bar\cQ^-_\a+\bar\l^A_R\,\bar R
 \end{align}
The variation at infinity constrains some parameters and also fixes the variation of various fields appearing in the asymptotic gauge fields. Below, we present the relations in the (2,0) sector, where fields and parameters are only function of $x_+$:
\begin{align*}
\chi^0&=-\cY^\prime+\frac{r}{l}\,\cY
\nonumber\\
\chi^-&=\tfrac12\,\cY^{\prime\prime}-\frac{r}{2\,l}\,\cY^\prime+\big(\frac{r^2}{4\,l^2}-\tfrac12\,\mathfrak L_+\big)\cY-\tfrac14\,(\psi_+\,\epsilon_--\psi_-\,\epsilon_+)
\nonumber\\
\epsilon^-_+&=-\epsilon_+^\prime+\frac{r}{2\,l}\,\epsilon_+-\tfrac12\,\psi_+\,\cY+\tfrac12{\rm i}\,\phi^A_R\,\epsilon_+
\nonumber\\
\epsilon^-_-&=-\epsilon_-^\prime+\frac{r}{2\,l}\,\epsilon_-+\tfrac12\,\psi_-\,\cY-\tfrac12{\rm i}\,\phi^A_R\,\epsilon_-
\end{align*}
where we called $\chi^+=\cY$, $\epsilon^+_+=\epsilon_+$ and $\epsilon^+_-=\epsilon_-$. The variations read:
\begin{align*}
\delta \mathfrak L_+&=-\cY^{\prime\prime\prime}+2\,\mathfrak L_+\,\cY^\prime+\mathfrak L_+^\prime\,\cY+\tfrac12\Big(\psi_+^\prime\,\epsilon_-+3 \,\psi_+\,\epsilon_-^\prime\Big)
-\tfrac12 \Big(\psi_-^\prime\,\epsilon_++3\,\psi_-\,\epsilon_+^\prime\,\Big)
\nonumber\\
&\;\;+\tfrac12{\rm i}\Big(\psi_+\,\epsilon_-\,^A\phi_R\,+\psi_-\,\epsilon_+\,\phi^A_R\,\Big)
\nonumber\\
\delta\psi_+&=2\,\epsilon_+^{\prime\prime}+\psi_+^\prime\,\cY+\tfrac32\,\psi_+\,\cY^\prime-{\rm i}\Big(\phi_R^{A\prime}\,\epsilon_++2\,\phi^A_R\,\epsilon_+^\prime\Big)-\mathfrak L_+\,\epsilon_+-\tfrac12{\rm i}\,\psi_+\,\phi^A_R\,\cY-\tfrac12\,\l^A_R\,\psi_+
\nonumber\\
&\quad-\tfrac12\,\phi^A_R\,\phi^A_R\,\epsilon_+
\nonumber\\
\delta\psi_-&=-2\,\epsilon_-^{\prime\prime}+\psi_-^\prime\,\cY+\tfrac32\,\psi_-\,\cY^\prime-
{\rm i}\Big(\phi_R^{A\prime}\,\epsilon_-+2\,\phi^A_R\,\epsilon_-^\prime\Big)+\mathfrak L_+\,\epsilon_-+\tfrac12{\rm i}\,\psi_-\,\phi_R\,\cY+\tfrac12\,\l^A_R\,\psi_-
\nonumber\\
&\quad+\tfrac12\,\phi^A_R\,\phi^A_R\,\epsilon_-
\nonumber\\
\delta\phi^A_R&={\rm i}\,\l_R^{A\prime}-\tfrac12\,{\rm i}\psi_+\,\epsilon_-\,-\tfrac12\,{\rm i}\psi_-\,\epsilon_+\,
\end{align*}
Now following the same procedure as  before, the asymptotic symmetry algebra 
for $(2,0)$ asymptotically AdS supergravity theory is straightforwardly  found. The non-trivial supertrace elements are:
\begin{equation}
\langle L_n , L_m \rangle =\tfrac12 \g_{nm}\;,\qquad \langle \cQ^\pm_\a , \cQ^\mp_\b \rangle = C_{\a\b}\;,\qquad \langle R , R\rangle =-\tfrac12\;,
\end{equation}
from which the generic charge reads
\begin{align}
Q[\cY,\epsilon_\mp,\lambda_R]&=-\frac{k_l}{4\,\pi}\,\int \mathfrak L_+\,\cY +\psi_+ \,\epsilon_--\psi_-\,\epsilon_++{\rm i}\phi^A_R\,\l^A_R
\nonumber\\
&=-\frac{2}{k_l}\sum_{n}\mathfrak{L}^+_n\,\cY_{-n}+\psi^+_n\epsilon^-_{-n}-\psi^-_n\epsilon^+_{-n}+{\rm i} R_n\l^A_{-n}
\end{align}
with $\epsilon^\pm_n$ modes of  $\epsilon_\pm$.
 The non trivial Poisson brackets are given as:
\begin{align}\label{pb1}
{\rm i}\{\mathfrak{L}^+_n,\mathfrak{L}^+_m\}_{PB}&=(n-m)\mathfrak{L}^+_{n+m}+\frac{c}{12}\,n^3\,\delta_{n+m,0}
\nonumber\\
{\rm i}\{R_n,R_m\}_{PB}&=\frac{k_l}{2}\,n\,\delta_{m+n,0}=\frac{c}{12}\,n\,\delta_{m+n,0}
\nonumber\\
{\rm i}\{\mathfrak L^+_n,\psi^\pm_\a\}_{PB}&=\Big(\frac{n}{2}-\a\Big)\,\psi^\pm_{\a+n}\mp
\frac{{1}}{2}[\Psi^\pm\,R]_{n+\a}
\nonumber\\
{\rm i}\{R_n,\psi^\pm_\a\}_{PB}&=\pm \tfrac12\,\psi^\pm_{\a+n}
\nonumber\\
\{\psi^+_\a,\psi^-_\b\}_{PB}&=\mathfrak L^+_{\a+\b}+(\a-\b)\,R_{\a+\b}+\tfrac12\,[R\,R]_{\a+\b}+
\frac{c}{6}\,\a^2\,\delta_{\a+\b}.
\end{align}
where the modes are defined as follows.:
\begin{align}\label{adsM1}
\mathfrak{L}^+_n&=\frac{k_l}{4\pi}\int d\varphi e^{in\varphi}\mathfrak{L}_+, \quad R_n= \frac{k_l}{4\pi}\int d\varphi e^{in\varphi}\phi^A_R,   \nonumber \\
\psi^{\pm}_{\alpha}&=\frac{k_l}{4\pi}\int d\varphi \psi^{\pm} e^{i\alpha\varphi}, \quad [\psi^{\pm}
R]_{\alpha}=\frac{k_l}{4\pi}\int d\varphi e^{i\alpha\varphi}\psi^{\pm}\phi^A_R,  \nonumber \\ 
&[RR]_{\alpha}=\frac{k_l}{4\pi}
\int d\varphi e^{i\alpha \varphi}\phi^A_R\phi^A_R 
\end{align}
Now, we need to redefine the generator $\mathfrak{L}_n$ by adding a term bilinear in the $\mathcal{R}$-current:  
\begin{equation}
\mathfrak{L}_n \rightarrow \hat{\mathfrak{L}}_n =\mathfrak{L}_n+\frac{1}{2}
(RR)_n
\end{equation}
This is a Sugawara shift on the Stress-tensor in presence of internal currents. The effect of this shift is shown below, where we write down the quantum (anti)commutator for the  theory using the same  convention as \ref{sec:BMSalgebra} :
\begin{align}\label{adsasympF}
[\hat{\mathfrak{L}}^+_n,\hat{\mathfrak{L}}^+_m]&=(n-m)
\hat{\mathfrak{L}}^+_{n+m}+\frac{c}{12}n^3\delta_{n+m,0}, \quad [R_n,\psi^{\pm}_{\alpha}]=
\pm\frac{1}{2}\psi^{\pm}_{n+\alpha}  \nonumber  \\
\{\psi^+_{\alpha},\psi^-_{\beta}\}&=\hat{\mathfrak{L}}^+_{\alpha+\beta}+(\alpha-\beta)
R_{\alpha+\beta}+
\frac{c}{6}\alpha^2\delta_{\alpha+\beta,0}  \\
[\hat{\mathfrak{L}}^+_n,R_m]&=-m\,R_{n+m}, \quad [\hat{\mathfrak{L}}^+_n,
\psi^{\pm}_{\alpha}]=\left(\frac{n}{2}-\alpha\right)\psi^{\pm}_{n+\alpha}, \quad
[R_n,R_m]=\frac{c}{12}n\,\delta_{n+m,0}  \nonumber
\end{align}
One can carry on similar computation for the $(0,2)$ sector and in this case, 
the constraints are:
\begin{align*}
\bar\chi^0&=\bar \cY^\prime-\frac{r}{l}\,\cY
\nonumber\\
\bar\chi^+&=\tfrac12\,\bar\cY^{\prime\prime}-\frac{r}{2\,l}\,\bar\cY^\prime+\big(\frac{r^2}{4\,l^2}-\tfrac12\,\mathfrak L_-\big)\bar\cY+\tfrac14\,(\bar\psi_+\,\bar\epsilon_--\bar\psi_-\,\bar\epsilon_+)
\nonumber\\
\bar\epsilon^+_+&=\bar\epsilon_+^\prime-\frac{r}{2\,l}\,\bar\epsilon_+-\tfrac12\,\bar\psi_+\,\bar 
\cY-\tfrac12\,{\rm i}\bar\phi^A_R\,\bar\epsilon_+
\nonumber\\
\bar\epsilon^+_-&=\bar\epsilon_-^\prime-\frac{r}{2\,l}\,\bar\epsilon_-+\tfrac12\,\bar\psi_-\,\bar 
\cY+\tfrac12{\rm i}\,\bar\phi^A_R\,\bar\epsilon_-
\end{align*}
where we called $\bar\chi^-=\bar\cY$, $\bar\epsilon^{\,-}_i=\bar\epsilon_i$. The variations 
read:
\begin{align*}
\delta \mathfrak L_-&=-\bar \cY^{\prime\prime\prime}+2\,\mathfrak L_-\,\bar \cY^\prime+\mathfrak L_-^\prime\,\bar \cY-\tfrac12\Big(\bar\psi_+^\prime\,\bar\epsilon_-+3 \,\bar\psi_+\,\bar\epsilon_-^\prime\Big)
+\tfrac12\Big(\bar\psi_-^\prime\,\bar\epsilon_++3\,\bar\psi_-\,\bar\epsilon_+^\prime\,\Big)
\nonumber\\
&\;\;-\tfrac12\,{\rm i}\Big(\bar\psi_+\,\bar\epsilon_-\,\bar\phi^A_R\,+\bar\psi_-\,
\bar\epsilon_+\,\bar\phi^A_R\,\Big)
\nonumber\\
\delta\bar\psi_+&=-2\,\bar\epsilon_+^{\prime\prime}+\bar\psi_+^\prime\,\bar \cY+\tfrac32\,\bar\psi_+
\,\bar \cY^\prime+{\rm i}\Big(\bar\phi_R^{A\prime}\,\bar\epsilon_++2\,\bar\phi^A_R\,
\bar\epsilon_+^\prime\Big)+\mathfrak L_-\,\bar\epsilon_+-\tfrac12\,{\rm i}\bar\psi_+\,\bar\phi^A_R\, \bar \cY-\tfrac12\,\bar\l^A_R\,\bar\psi_+
\nonumber\\
&\quad+\tfrac12\,\bar\phi^A_R\,\bar\phi^A_R\,\bar\epsilon_+
\nonumber\\
\delta\bar\psi_-&=2\,\bar\epsilon_-^{\prime\prime}+\bar\psi_-^\prime\,\bar \cY+\tfrac32\,\bar\psi_-
\,\bar \cY^\prime+{\rm i}\Big(\bar\phi_R^{A\prime}\,\bar\epsilon_-+2\,\bar\phi^A_R\,
\bar\epsilon_-^\prime\Big)-\mathfrak L_-\,\bar\epsilon_-+\tfrac12\,{\rm i}\bar\psi_-\,\bar\phi^A_R\,\bar \cY+\tfrac12\,\bar\l^A_R\,\bar\psi_-
\nonumber\\
&\quad-\tfrac12\,\bar\phi^A_R\,\bar\phi^A_R\,\bar\epsilon_-
\nonumber\\
\delta\bar\phi^A_R&={\rm i}\bar\l_R^{A\prime}+\tfrac12\,{\rm i}\bar\psi_-\bar\epsilon_-+\tfrac12\,{\rm i}\bar\psi_-\,\bar\epsilon_+
\end{align*}
The  non-zero supertraces elements are:
\begin{equation}
\langle \bar L_n , \bar L_m \rangle =-\tfrac12\g_{nm}\;,\qquad \langle \bar \cQ^\pm_\a , \bar \cQ^\mp_\b \rangle = - C_{\a\b}\;,\qquad \langle \bar R , \bar R\rangle =\tfrac12\;,
\end{equation}
and the charge of the barred sector reads
\begin{align}
\bar Q[\bar \cY,\bar \epsilon_\mp,\bar \lambda_R]&=-\frac{k_l}{4\,\pi}\,\int \mathfrak L_-\,\bar \cY -\bar \psi_+ \,\bar \epsilon_-+\bar \psi_-\,\bar \epsilon_++{\rm i}\bar \phi^A_R\,\bar \l^A_R
\nonumber\\
&=-\frac{2}{k_l}\sum_{n}\mathfrak{L}^-_n\,\bar \cY_{-n}-\bar \psi^+_n\bar \epsilon^-_{-n}+\bar \psi^-_n\bar \epsilon^+_{-n}+{\rm i} \bar  R_n\bar \l^A_{-n}
\end{align}
with $\bar\epsilon^\pm_n$ modes of  $\bar\epsilon_\pm$.
Finally, the asymptotic form of the Poisson brackets between various modes take identical form of the $(2,0)$ 
sector as,
\begin{align}\label{pb2}
{\rm i}\{{\mathfrak L}^-_n,{\mathfrak L}^-_m\}_{PB}&=(n-m)\,
{\mathfrak L}^-_{n+m}+\frac{\bar c}{12}\,n^3\,\delta_{m+n,0}
\nonumber\\
{\rm i}\{\bar R_n,\bar R_m\}_{PB}&=\frac{k_l}{2}\,n\,\delta_{m+n,0}=\frac{\bar c}{12}\,n\,\delta_{m+n,0}
\nonumber\\
{\rm i}\{{\mathfrak L}^-_n,\bar\psi^\pm_\a\}_{PB}&=\Big(\frac{n}{2}-\a\Big)\,\bar\psi^\mp_{\a+n}\pm
\frac{{\rm 1}}{2}[\bar\Psi^\pm\,\bar R]_{n+\a}
\nonumber\\
{\rm i}\{\bar R_n,\bar\psi^\pm_\a\}_{PB}&=\pm \tfrac12\,\bar\psi^\pm_{\a+n}
\nonumber\\
\{\bar\psi^+_\a,\bar\psi^-_\b\}_{PB}&={\mathfrak L}^-_{\a+\b}+\,(\a-\b)\,\bar R_{\a+\b}+\tfrac12\,[\bar R\,\bar R]_{\a+\b}+\frac{c}{6}\,\a^2\,\delta_{\a+\b}
\end{align}
where the modes are defined as follows.:
\begin{align}\label{adsM2}
{\mathfrak{L}}^-_n&=\frac{k_l}{4\pi}\int d\varphi e^{-in\varphi}\mathfrak{L}_-,\quad \bar{ R }_n=
\frac{k_l}{4\pi}\int d\varphi e^{-in\varphi}\bar\phi^A_R,   \nonumber \\
\bar\psi^{\pm}_{\alpha}&=\frac{k_l}{4\pi}\int d\varphi \bar\psi^{\pm} e^{-i\alpha\varphi}, \quad 
[\bar\psi^{\pm}
\bar{ R }]_{\alpha}=\frac{k_l}{4\pi}\int d\varphi e^{-i\alpha\varphi}\bar\psi^{\pm}\bar\phi^A_R,  \nonumber \\ 
&[\bar R\bar R]_{\alpha}=\frac{k_l}{4\pi}\int d\varphi e^{-i\alpha \varphi}
\bar\phi^A_R\bar\phi^A_R  
\end{align}
Notice that the definition of Fourier transform in barred and unbarred sectors are different. This  is ultimately due to the fact  that the two sectors depend exclusively on $x^-$ and $x^+$ respectively, so  that one can expand all the arguments in power series of $1/l$, and the fields and  charges in the barred  sector will depend on $-\varphi$. 
Finally using the same convention for writing the suitable quantum commutators in the 
barred sector, the asymptotic symmetry algebras for the generators of the barred sector, 
i.e. of $(0,2)$ three dimensional AdS theory takes exactly
identical form as the one for $(2,0)$ 
three dimensional AdS theories presented in \ref{adsasympF}. Here also, we required 
a Sugawara shift of the Stress-tensor as 
\begin{equation}
{\mathfrak{L}}^-_n \rightarrow \hat{\mathfrak{L}}^-_n =\mathfrak{L}^-_n+\frac{1}{2}
(\bar R \bar R)_n
\end{equation} to get the final form of the algebra which is identical to \eqref{adsasympF}.

Note that we started with identical copies of bulk symmetry algebras for 
$(2,0)$ and $(0,2)$ sectors as given in appendix \ref{appb} and, as  a consequence, the asymptotic 
algebras of the modes of 
the conserved charges are also identical, differences in the sign of  the Fourier modes  notwithstanding. It was shown  in \cite{Banerjee1} that by properly combining the two algebras, one immediately obtains the modes of \eqref{bmsmodes} and the corresponding BMS algebra \eqref{flatalgebra}.

\subsection{$N=4$ super-BMS$_3$ from $N=(2,2)$ super-AdS$_3$}
In this section, we shall explicitly show the relations between the generators, gauge fields  components and  gauge parameters of two copies of the super conformal algebras and the flat $N=4$ BMS$_3$ algebra. As it is  easy to understand, in fact, the latter quantities can
be obtained from the linear combinations of the former ones. Before doing so, let us recall that a In\"on\"u-Wigner contraction of two 
copies of Super-conformal algebra gives us the Super-Poincare algebra. The contraction is 
defined in the large AdS radius limit $l \rightarrow \infty$.  The
level of the corresponding Chern-Simons actions are related as $k_l= k \cdot
l$. The generators of the flat algebra can be obtained from the AdS ones as, 
\begin{align*}
\cL_n&=L_n-\bar L_{-n},\quad M_n=\frac{L_n+\bar L_{-n}}{l}, \quad \cR=R-\bar R,\quad \cS=\frac{R+\bar R}{l}\nonumber\\
q^{1\,\pm}_\a&=\sqrt{\frac{2}{l}}\cQ^{\pm}_\a\;,\qquad q^{2\,\pm}_\a=\sqrt{\frac{2}{l}}\bar\cQ^{\pm}_{-\a}
\end{align*}
It is easy to check that the asymptotic gauge field  and the gauge transformation parameter of the flat theory is obtained from the AdS ones in the limit $l \rightarrow \infty$ as, 
\begin{equation}
\cA=A_++A_{-}, \quad \L=\L_++\L_{-}.
\end{equation}
which can be immediately decomposed in the sum of the gauge  variations of the two superconformal sectors, up the  remembering that the unbarred and barred sectors are functions of $x^+$ and $x^-$ coordinate respectively. We further need to use the following maps of various charges, whose algebra is indeed the asymptotic super-BMS$_3$ algebra we have derived before \footnote{Again, when writing down the BMS charges in terms of Virasoro modes, the presence of the Chern-Simons level is crucial to  obtain the correct scaling.}:
\begin{align}\label{relationF}
 \cM&=\mathfrak L_+ + \mathfrak L_-\;,\qquad \cN =l\,(\mathfrak L_+-\mathfrak L_-)
,\quad \phi=l(\phi^A_R-\bar \phi^A_R),
\nonumber\\
\rho&=\phi^A_R+\bar\phi^A_R, \quad\psi^{\a}_\pm=\frac{1}{\sqrt{2l}}
\Psi^{1\,\a}_{\pm}\;,\qquad \bar\psi^{-\a}_\pm=\frac{1}{\sqrt{2l}}\Psi^{2\,\a}_{\pm}\;,
\end{align}
for the parameters:
\begin{align*}
\epsilon_{\pm}^\a&=\sqrt{\frac{2}{l}}\,\z^{1\,\a}_{\pm}\;,\qquad 
\bar\epsilon_{\pm}^{-\a}=\sqrt{\frac{2}{l}}\,\z^{2\,\a}_{\pm}\;,\qquad \U^n=\frac{\chi^n-\bar\chi^{-n}}{2},
\nonumber\\
\xi^n&=l\frac{\chi^n+\bar\chi^{-n}}{2}, \quad \l_\cR=\frac{\l^A_R-\bar\l^A_R}{2}\;,\qquad \l_\cS=l\frac{\l^A_R+\bar\l^A_R}{2}
\end{align*}
%The constraint relations and the variation of various fields in both theories also follow directly by noticing that,
%\begin{align}\label{RELATION}
%\delta\cA_\varphi&=\delta (A_++A_-) \nonumber \\
%\partial_\varphi\L+[\cA_\varphi,\L]&= \partial_+\Lambda_++[A_+,\Lambda_+]-\big(\partial_-\Lambda_-+[A_-,\Lambda_-]\big)
%\end{align}
To obtain  \eqref{eq:ident_1}-\eqref{eq:ident_2} from \eqref{eq:constr_2scsectors} one needs to make use of the change of variables identity: 
\begin{equation}
\partial_\varphi=\partial_+-\partial_-\;,\qquad \partial_u=\frac{\partial_++\partial_-}{l}.
\end{equation}
and similarly for the constraints on the parameters. 
%Equation \ref{RELATION} is one of the main results of this paper and is an extremely 
%useful one. It tells us, given the knowledge of the asymptotic gauge fields  and 
%gauge transformation parameters of the two sectors of the AdS theory, how one can 
%construct the corresponding ones for the  asymptotic flat theory. In a related
%work \cite{BLN3}, we have used this relation to find the $N=8$ Super BMS$_3$ algebra.\\ 
Finally, as proposed in \cite{BLN1}, the three dimensional $N=4$ BMS algebra  
\ref{flatalgebra} can 
be obtained by two identical copies of asymptotic $(2,0)$ and $(0,2)$ AdS$_3$ algebras with the following identification for the charges and their Fourier modes 
\begin{align}
\label{eq:contraction_1bis}
\mathfrak{J}_m&=\lim_{\epsilon\rightarrow 0}\,(\mathfrak{L}^+_m-\mathfrak{L}^-_{-m})\;, &                                                    \hat{\cM}_m&=\lim_{\epsilon\rightarrow 0}\,\epsilon (\mathfrak{L}^+_m+\mathfrak{L}^-_{-m})\;,
\nonumber\\
\Psi^{1,\pm}_r&=\lim_{\epsilon\rightarrow 0}\,\sqrt{\epsilon}\,\psi^{\pm}_{r}\;,
&
\Psi^{2,\pm}_r&=\lim_{\epsilon\rightarrow 0}\,\sqrt{\epsilon}\,\bar\psi^{\pm}_{-r}\;,
\nonumber\\
c_J&=\lim_{\epsilon\rightarrow 0}(c-\cbar)\;,&
c_M&=\lim_{\epsilon\rightarrow 0}\,\epsilon\,(c+\bar c)\;,
\nonumber\\
\cR_m&=\lim_{\epsilon\rightarrow 0}\, (R_m-\bar R_{-m})\;, &
                                                       \;\cS_m&=\lim_{\epsilon\rightarrow 0}\,\epsilon \,(R_m+\bar R_{-m})\;,
\end{align}
where $\e= \frac1l$. The above identification follows directly from relation \ref{relationF} and
definitions of various modes as given in \ref{bmsmodes}, \ref{adsM1} and \ref{adsM2}.

Finally we end this section by justifying the Sugawara shifts on the two flat algebra generators $\mathfrak{J}$ and $\cM$ to 
obtain the algebra \ref{flatalgebra}.
% While the first one is motivated by the Sugawara shift of the Stress-energy tensor, in presence of a $R-$ symmetry
%current,  the later was proposed to get proper energy bound.
If we think of the BMS$_3$ algebra as a limit of two copies of AdS$_3$ algebras,
then it is obvious to realize why both $\mathfrak{J}$ and  $\cM$ require a shift. Writing those shifts in terms of fields we have:
\begin{equation}
\hat{\mathfrak{L}}_+=\mathfrak{L}_++\tfrac12 \phi_A^2\;,\quad \hat{\mathfrak{L}}_-=
\mathfrak{L}_-+\tfrac12 \bar\phi_A^2\;,
\end{equation}
It is easy to check that, the BMS$_3$ charges $\mathfrak{J}$ and $\cM$, which are 
combinations of the two above AdS$_3$ charges $\mathfrak{L}_\pm$ will pick up certain shifts. In 
particular the shift in $\cM$ comes out as,
\begin{align*}
\cM&= (\mathfrak{L}_++\mathfrak{L}_-)=(\hat{\mathfrak{L}}_+
+\hat{\mathfrak{L}}_-)-\tfrac12(\phi_A^2+\bar\phi_A^2)
=\hat{\cM}-\tfrac14(\rho^2+(\phi/l)^2) \nonumber 
%\hat{\cM}&=\cM+\tfrac14\phi\phi
\end{align*}
where we used the definitions of the  R- and S-symmetry gauge fields. Similarly, one obtains the shift for $\mathfrak{J}$ (more  care needs to be exercised in that case,where it is crucial to expand the Virasoro R-symmetry fields in powers of $1/l$). In the limit $l\to\infty$, we finally get the shifts (in terms of the modes) as :
\begin{equation}
\hat{\mathfrak{J}}_n=\mathfrak{J}_n+\tfrac12 [\cR\cS]_n \;,\qquad 
\hat{\cM}_n=\cM_n+\tfrac14 [\cS\cS]_n 
\end{equation}
These  are indeed the  correct Sugawara shifts for the BMS$_3$ generators that simplify 
the algebra  notably. The most important simplification happens at the level of the 
anti-commutator  of the supercharges, as the non-linear term $[\cS\cS]$ is 
immediately absorbed  inside $\cM$. 

\vspace{1cm}
{\bf Acknowledgements}\\

\noindent
We would like to thank Rudranil Basu, Dileep Jatkar, Wout Merbis and Sunil Mukhi for useful discussions.  We would also like to thank Nemani Suryanarayana for bringing an important reference to our notice. Our work is partially supported by the following
Government of India Fellowships/Grants: N.B. and I.L. by a Ramanujan Fellowship, D.S.T. and T.N.
by a UGC Fellowship. We thank the people of India for their generous support for the
basic sciences.

\appendix
\section{Conventions}\label{appa}
In this paper we follow conventions similar to \cite{Lodato}. We will list them here to maintain the paper self-contained. \\
The antisymmetric Levi-Civita symbol has component $\epsilon_{012} = +1$ and the tangent space metric is the 3D Minkowski metric
\begin{equation}
\eta_{ab} = \left( \begin{array}{ccc} -1 & 0 & 0 \\ 0 & 1 & 0 \\ 0 & 0 & 1\end{array} \right)
\end{equation}
The $\Gamma$-matrices satisfying the three dimensional Clifford algebra $\{\Gamma_a, \Gamma_b\} =2 \eta_{ab}$ are:
\begin{equation}
\Gamma_0 =  i \sigma_2 \,, \qquad \Gamma_1 = \sigma_1 \,, \qquad \Gamma_2 = \sigma_3\,,
\end{equation}
with $\sigma_i$ the Pauli matrices:
\begin{equation}
\sigma_1 = \left(\begin{array}{cc} 0 & 1 \\ 1 & 0\end{array} \right)\,, \qquad 
\sigma_2 = \left(\begin{array}{cc} 0 & -i \\ i & 0\end{array} \right)\,, \qquad
\sigma_1 = \left(\begin{array}{cc} 1 & 0 \\ 0 & -1\end{array} \right)\,.
\end{equation}
Finally, the charge conjugation matrix $C = i\sigma_2$, or explicitly
\begin{equation}
C_{\alpha\beta} = \varepsilon_{\alpha\beta} = C^{\alpha\beta}= \left(\begin{array}{cc} 0 & 1 \\ -1 & 0\end{array} \right)\,.
\end{equation}
Throughout this paper the fermionic indices $\a,\b$ run over $-,+$ (contrarily to \cite{Lodato} where they run over $+,-$). The supercharges are also taken to be Grassmann quantities, as are the fermion parameters and the gravitini.\\ All spinors in this work are Majorana and the Majorana conjugate of a spinor $\psi^{\alpha}$ is $\bar{\psi}_{\alpha} = C_{\alpha\beta}\psi^{\beta}$. Our conventions imply that we can use the identities
\begin{align}
\Gamma_a\Gamma_b & = \epsilon_{abc}\Gamma^c + \eta_{ab} \mathbb{I}\,, &&& \Gamma^a{}^{\alpha}{}_{\beta} \Gamma_a{}^{\gamma}{}_{\delta} & = 2 \delta^{\alpha}_{\delta} \delta^{\gamma}_{\beta} - \delta^{\alpha}_{\beta}\delta^{\gamma}_{\delta}\,, \\
C^T & = - C\,, &&& C \Gamma_a & = - (\Gamma_a)^T C
\end{align}
In verifying the closure of the supersymmetry algebra on the fields and the off-shell invariance of the action, the three dimensional Fierz relation is useful.
\begin{equation}\label{Fierz}
\zeta\bar{\eta} = - \frac12 \bar{\eta}\, \zeta \, \mathbb{I} - \frac12 (\bar{\eta}\Gamma^a \zeta)\Gamma_a\;,
\end{equation}
Other useful identities are:
\begin{align*}
\bar\psi \G_a\,\eta&=\bar\eta\,\G_a\,\psi
\nonumber\\
\bar\psi \G_a\,\epsilon&= -\bar\epsilon\,\G_a\,\psi
\end{align*}
where $\psi,\eta$ are Grassmannian one-forms, while 
$\epsilon$ is a Grassmann paramter.
\noindent It is sometimes convenient to change basis of the tangent space to one more suited for the $isl(2)$  algebra in the bosonic sector of flat space supergravity. We do this by choosing a map to bring the generators of $SO(2,1)$ ($[J_a,J_b] = \epsilon_{abc}J^c$) to those of $SL(2,\bR)$ satisfying $[L_n,L_m] = (n-m)L_{n+m}$. This defines a matrix $U^a{}_n$ as a map from the tangent space metric $\eta_{ab}$ with $a,b=\{0,1,2\}$ to the metric $\gamma_{nm}$ defined in \eqref{gammadef} with $n,m= \{-1,0,+1\}$, satisfying
\begin{equation}
L_n = J_a\, U^a{}_n \,.
\end{equation}
An explicit representation of $U^a{}_n$ that does the job is for instance
\begin{equation}\label{Umat}
U^a{}_n = \left( \begin{array}{ccc} -1 & 0 & -1 \\ -1 & 0 & 1 \\ 0 & 1 & 0 \end{array} \right)\,.
\end{equation}
In this basis the gamma matrices satisfy a Clifford algebra with
\begin{equation}
\label{Cliffnm}
\{\tilde{\Gamma}_m, \tilde{\Gamma}_n\} = 2 \gamma_{nm} \equiv 2 \left( \begin{array}{ccc} 0 & 0 & -2 \\ 0 & 1 & 0 \\ -2 & 0 & 0 \end{array}\right) \qquad \text{with: } n,m= -1 , 0, +1\,.
\end{equation}
A real representation for the gamma matrices with $n,m$ indices can be obtained by taking $\tilde{\Gamma}_n= U^a{}_n \Gamma_a$, or explicitly:
\begin{align}
\label{gammadef}
\tilde{\Gamma}_{-1} & =  - (\sigma_1 + i \sigma_2) = \left(\begin{array}{cc} 0 & -2 \\ 0 & 0\end{array} \right)\,, \\ 
\tilde{\Gamma}_0 & = \sigma_3 = \left(\begin{array}{cc} 1 & 0 \\ 0 & -1\end{array} \right)\,, \\
\tilde{\Gamma}_{+1} & = \sigma_1 - i \sigma_2 = \left(\begin{array}{cc} 0 & 0 \\ 2 & 0\end{array} \right)\,.
\end{align}
In addition to the Clifford algebra \eqref{Cliffnm}, the gamma matrices now satisfy the commutation relations
\begin{equation}\label{Gammacom}
[\tilde{\Gamma}_{n} , \tilde{\Gamma}_{m}] = 2(n-m)\tilde{\Gamma}_{n+m}\,,
\end{equation}
which is the $sl(2,\mathbb{R})$ algebra.
\section{Construction of the supertrace elements}\label{appc}
In this appendix, we shall outline the procedure to obtain the supertrace elements for a given algebra. Below, we present the computation for (2,0) AdS algebra, that is presented in the last appendix.
Super trace element is computed from non-degenerate bilinear form of a given algebra.  For this, we construct a quadratic scalar combination of all the generators and impose that it commutes with all the generators, so that it is a Casimir operator. The construction of this quadratic scalar invariant is quite easy. Let us focus on the (2,0) algebra first, and find its non-zero supertrace elements. 
Now let us start with the most generic possible bilinear form $W$ :
\begin{equation}
W= a\,\eta^{ab}\,J_a\,J_b+b\,C^{\a\b}\,Q^+_\a\,Q^-_\b+ \bar b\,C^{\a\b}\,Q^-_\a\,Q^+_\b+
c\,C^{\a\b}\,Q^+_\a\,Q^+_\b+\bar c\,C^{\a\b}\,Q^-_\a\,Q^-_\b+d\,R\,R
\end{equation}
By demanding that W commutes with all the generators of the (2,0) super algebra, we can fix the factors $(a,b,\bar b,c,\bar c,d)$. In this process, we need to make sure that the final Casimir is non-degenerate.  We will use the identities:
\begin{align}
[A\,B,C]&=A[B,C]+[A,C]B \;,\qquad [A\,B,C]=A\{B,C\}-\{A,C\}B\;, \nonumber \\
C_{\a\b}\,C^{\b\g}&=\d_\a^\g, \quad C\G=(C\G)^T, \quad  \G C=(\G C)^T.
\end{align}
The parameters $(a,b,\bar b,c,\bar c,d)$ get fixed as,
\begin{equation}
a=b=\bar b= -d\;,\qquad c=\bar c=0\;.
\end{equation}
So overall, the invariant reads: 
\begin{equation}
W=a(\eta^{ab}\,J_a\,J_b+C^{\a\b}\,Q^+_{\a}\,Q^{-}_{\b}+C^{\a\b}\,Q^{-}_{\a}\,Q^{+}_{\b}-\,R\,R)
\end{equation}
From $W$ we extract all the supertrace elements (see also book by Blagojevic M. "Gravitation and gauge symmetries", Appendix L), by taking the inverse of the matrices $\eta^{ab}$, $C^{\a\b}$ and $\mathbb{I}$:
\begin{equation}
\label{eq:inv_trace_el}
\langle J_a , J_b \rangle =\frac{1}{a}\, \eta_{ab}\;,\qquad \langle Q^+_\a , Q^-_\b \rangle =\langle Q^-_\a , Q^+_\b \rangle= \frac{1}{a}\, C_{\a\b}\;,\qquad \langle R , R \rangle =-\frac{1}{a}
\end{equation}
Similarly for the (0,2) sector, the supertrace element is given as, 
\begin{equation}
\label{eq:stJbarJ}
\langle \bar J_a , \bar J_b \rangle=\frac{1}{\bar a}\,\eta_{ab}\;,\qquad \langle \bar Q^{\pm}_\a , \bar Q^{\mp}_\b\rangle =\frac{1}{\bar a}\, C_{\a\b} \;,\qquad  \langle \bar R , \bar R \rangle = -\frac{1}{\bar a}
\end{equation}
where we have kept the overall factors $a,d$ in the supertrace of both the sectors, because they correspond to an overall normalization in the action. These constant factors get fixed to $a=-\bar  a=-2$ for the bosonic action to contain the Einstein-Hilbert term.
The super Poincare generators $\cJ_a$, $P_a$, $\cQ^{1,2\;,\pm}_r$, $\cR$ and $\cS$ are given in terms of the super conformal generators as given in appendix \ref{appd}. Hence, we can find the supertrace elements for flat generators as linear combinations of the AdS$_3$ supertrace-elements
(the factor of $1/l$ is absorbed in the Chern-Simons level of the action, hence neglected below):
\begin{align}
\langle \cJ_a , P_b \rangle&=\langle (J_a+\bar J_a),(J_a-\bar J_a) \rangle=\eta_{ab}\;,
\nonumber\\
 \langle \cQ^{1\pm}_\a , \cQ^{1\mp}_\b\rangle&=(\sqrt{2})^2 \langle \,Q^{\pm}_\a  , Q^{\mp}_\b\rangle=C_{\a\b}\;,
 \nonumber\\
 \langle \cQ^{2\pm}_\a , \cQ^{2\mp}_\b\rangle&=(\sqrt{-2})^2 \langle \,\bar Q^{\pm}_\a  , \bar Q^{\mp}_\b\rangle=C_{\a\b}\;,
\nonumber\\ 
 \langle \cR , \cS \rangle&=\langle (R- \bar R),(R+\bar R)\rangle =-1.
\end{align}
(see Appendix \ref{appd} for the change of  basis of the generators).\\
When dealing with the asymptotic  algebra,the overall factor is necessary to obtain  the correct normalization of the charges.
\section{The (0,2) and (2,0) AdS sectors}\label{appb}
Below we present the $N=(2,0)$ and $(0,2)$ superconformal algebras, the global bulk algebras for the corresponding AdS supergravity theories.
\begin{align}
\label{eq:Virasoro20bis}
[J_a,J_b]&=\epsilon_{abc}\,J^c\;, \qquad\qquad  [J_a,R]=0\;, \qquad\quad   
[R,R]=0\;,\nonumber\\
 [J_a,Q^\pm_\a]&=\tfrac12\,(\G_a)^\b{}_\a\,Q^\pm_\b\;,\qquad\qquad \qquad \qquad \quad 
[R,Q^\pm_\a]=\pm \tfrac12 Q^\pm_\a\;,
\nonumber\\
\{Q^+_\a,Q^-_\b\}&=-\tfrac12\,(C\,\G^a)_{\a\b}\,J_a-\tfrac12\,C_{\a\b}\,R \;, 
\qquad \{Q^\pm_\a,Q^\pm_\b\}=0\;.
\end{align}
\begin{align}
\label{eq:Virasoro02bis}
[\bar J_a,\bar J_b]&=\epsilon_{abc}\,\bar J^c\;, \qquad\qquad  [\bar J_a,\bar R]=0\;, \qquad\quad   
[\bar R,\bar R]=0\;,\nonumber\\
 [\bar J_a,\bar Q^\pm_\a]&=\tfrac12\,(\G_a)^\b{}_\a\,\bar Q^\pm_\b\;,\qquad\qquad \qquad \qquad \quad 
[\bar R,\bar Q^\pm_\a]=\pm \tfrac12 \bar Q^\pm_\a\;,
\nonumber\\
\{\bar Q^+_\a,\bar Q^-_\b\}&=-\tfrac12\,(C\,\G^a)_{\a\b}\,\bar J_a-\tfrac12\,C_{\a\b}\,\bar R \;, 
\qquad \{\bar Q^\pm_\a,\bar Q^\pm_\b\}=0\;.
\end{align}
Here, $a,b = 0,1,2$ and $\alpha, \beta = \pm \frac{1}{2}$. Our convention for $(\G_a)^\b{}_\a$ and $\,C_{\a\b}$ are presented in the first appendix.
With gauge fields
\begin{equation}
A=A^a\,J_a+\sum_{i=\pm}\psi^\b_i\,Q^i_\b +\phi_R\,R\;,\qquad \bar A=\bar A^a\,\bar J_a+\sum_{i=\pm}\eta^\b_i\,\bar Q^i_\b+\bar \phi_R\,\bar R
\end{equation}
where $A^a=\o+\tfrac1{l} e^a$ and $\bar A^a=\o-\tfrac1{l} e^a$, one can build the the supersymmetric action:
\begin{align}
S=\frac{1}{16\,\pi\,G}\,\int&\Big[2\,e_a\,R^a+\tfrac2{l^2}\,e+\tfrac{l}2\,\bar\psi_-\,D\psi_++\tfrac{l}2\,\bar\psi_+\,D\psi_--\tfrac{l}2\,\bar\eta_-\,D\eta_+-\tfrac{l}2\,\bar\eta_+\,D\eta_-
\nonumber\\
&+\tfrac1{4}\big(\bar\psi_+\,e_a\,\G^a\,\psi_-+\bar\psi_-\,e_a\,\G^a\,\psi_+-\bar\eta_+\,e_a\,\G^a\,\eta_--\bar\eta_-\,e_a\,\G^a\,\eta_+\big)
\nonumber\\
&-\tfrac{l}2(\phi_R {\rm d} \phi_R-\bar\phi_R {\rm d} \bar\phi_R)\Big]
\end{align}
by using the  supertrace elements \eqref{eq:stJbarJ}. The covariant derivatives read:
\begin{align*}
D\psi_+&={\rm d}\psi_++\tfrac12\,\phi_R\psi_++\tfrac12\,\o\,\G\,\psi_+\;,\qquad D\psi_-={\rm d}\psi_--\tfrac12\,\phi_R\psi_-+\tfrac12\,\o\,\G\,\psi_-
\nonumber\\
D\eta_+&={\rm d}\eta_++\tfrac12\,\bar\phi_R\eta_++\tfrac12\,\o\,\G\,\eta_+\;,\qquad\;\;\, D\eta_-={\rm d}\eta_-+\tfrac12\,\bar\phi_R\eta_-+\tfrac12\,\o\,\G\,\eta_-
\end{align*}
To obtain the flat action \eqref{eq:flat_action}, we take the limit $l\rightarrow \infty$ combined with the following redefinitions for the fermions and R-symmetry generators :
\begin{equation}
\psi^\a_\pm\rightarrow\sqrt{\frac{2}{l}}\psi^{1\a}_\pm\;,\qquad \eta^\a_\pm\rightarrow\sqrt{-\frac{2}{l}}\psi^{2\a}_\pm\;,\qquad \phi_R\rightarrow (\frac{\sigma}{l}+\upsilon)\;,\qquad \bar\phi_R \rightarrow (\frac{\sigma}{l}-\upsilon)
\end{equation}

%The  action is invariant off-shell under the supersymmetry transformation laws $\delta A = {\rm d}\lambda + [A, \lambda]$ with $\lambda = \varepsilon_{\pm}^{\alpha} Q^{\pm}_{\alpha} + \vartheta_\pm^{\alpha} \bar Q^{\pm}_{\alpha}$. In terms of the fields these transformations read:
%
%\begin{align}
%\delta e_{\mu}{}^a & = - \frac12 \sum_{\beta=\pm}\left(\bar{\varepsilon}_\beta \Gamma^a \psi_{\mu\,-\beta} + \bar{\vartheta}_\beta \Gamma^a \eta_{\mu\,-\beta} \right) \, && \delta \omega_{\mu}{}^a  = 0 \,,\quad \delta\phi=0 \\
%\delta \psi^{1\a}_{\pm\mu} & = D_{\mu}\varepsilon^\a_\pm={\rm d}\varepsilon^\a_\pm+\tfrac12\,\omega^a\,(\G_a)^\a{}_\g\,\varepsilon_\pm^\g\pm \tilde\phi\,\varepsilon^\a_\pm \,,\\
% \delta \psi^{2\a}_{\pm\mu}  & = D_{\mu}\vartheta^\a_\pm={\rm d}\vartheta^\a_\pm+\tfrac12\,\omega^a\,(\G_a)^\a{}_\g\,\vartheta_\pm^\g\,\pm \tilde\phi\,\vartheta^\a_\pm\,\\
%\delta \tilde\rho&=-\tfrac14(\bar\psi^1_+\,\varepsilon_--\bar\psi^1_-\,\varepsilon_++\bar\psi^2_+\,\vartheta_--\bar\psi^2_-\,\vartheta_+)
%\end{align}

\section{Sum up all the notations and change of basis}\label{appd}
In this appendix, we sum up the notations for the two basis for AdS and Flat algebra that we have used in our computation. Although required relations are mentioned 
in the main draft, here we sum them up in a compact form for future reference. First we write down the notations for various generators ,fields and gauge transformation parameters in all four cases : 
\begin{center}
AdS supergravity
\begin{tabular}{|c||cccccc|}
\hline
\rule{0pt}{3ex}
generator  & $L_n$ & $\bar L_n$ & $\cQ^\pm_\a$ & $\bar\cQ^\pm_\a$  & $R$ & $\bar R$  \\%[.5mm] \hline
 gauge fields & $A^n$ & $\bar A^n$ & $\psi_\pm^{A\a}$ & $\bar\psi_\pm^{A\a}$ &  
$\phi^A_R$ & $\bar\phi^A_R$ \\%[.5mm] \hline
parameters  & $\chi^n$ & $\bar\chi^n$ & $\epsilon_\pm^\a$ &  $\bar\epsilon_\pm^\a$ & $\l^A_R$ & $\bar\l^A_R$  \\%[.5mm] \hline

\hline
\end{tabular}
\end{center}
\begin{center}
AdS supergravity
\begin{tabular}{|c||cccccc|}
\hline
\rule{0pt}{3ex}
generator  & $J_a$ & $\bar J_{a}$ & $Q^\pm_\a$ & $\bar Q^\pm_\a$  & $R$ & $\bar R$  \\%[.5mm] \hline
 gauge fields & $A^a$ & $\bar A^a$ & $\psi_\pm^\a$ & $\bar\psi_\pm^\a$ &  
$\phi_R$ & $\bar\phi_R$ \\%[.5mm] \hline
parameters  & $...$ & $...$ & $\varepsilon_\pm^\a$ &  $\vartheta_\pm^\a$ & $\l_R$ & $\bar\l_R$  \\%[.5mm] \hline
\hline
\end{tabular}
\end{center}
\begin{center}
Poincare' supergravity
\begin{tabular}{|c||cccccc|}
\hline
\rule{0pt}{3ex}
generator  & $\cJ_a$ & $P_{a}$ & $\cQ^{1\pm}_\a$ & $\cQ^{2\pm}_\a$  & $\cR$ & $\cS$  \\%[.5mm] \hline
 gauge fields & $\o^a$ & $e^a$ & $\psi^{1\a}_\pm$ & $\psi_\pm^{2\a}$ &  
$\tilde{\phi}$ & $\tilde\rho$ \\%[.5mm] \hline
parameters  & $...$ & $...$ & $\theta_\pm^{1\a}$ &  $\theta_\pm^{2\a}$ & $\l_\cR$ & $\bar\l_\cS$  \\%[.5mm] \hline
\hline
\end{tabular}
\end{center}
\begin{center}
Poincare' supergravity
\begin{tabular}{|c||cccccc|}
\hline
\rule{0pt}{3ex}
generator  & $\cL_n$ & $M_n$ & $q^{1\pm}_\a$ & $q^{2\pm}_\a$  & $\cR$ & $\cS$  \\%[.5mm] \hline
 gauge fields & $\o^n$ & $e^n$ & $\Psi^{1\a}_\pm$ & $\Psi_\pm^{2\a}$ &  
$\phi$ & $\rho$ \\%[.5mm] \hline
parameters  & $\U^n$ & $\xi^n$ & $\zeta_\pm^{1\a}$ &  $\zeta_\pm^{2\a}$ & $\l_\cR$ & $\bar\l_\cS$  \\%[.5mm] \hline
\hline
\end{tabular}
\end{center}
 Next we present the relations between the generators, fields and parameters for the above cases:
~~~~~~~~~~~~~~~~~\\
\textbf{Relation between the two AdS basis:}
\begin{align}
L_n&=J_a\,U^a_n\;,\quad \bar L_n=\bar J_a U^a_n\;,\quad \cQ^{\pm}_\a=\sqrt{2}Q^{\pm}_\a\;,\quad \bar\cQ^{\pm}_\a=\sqrt{2}\bar Q^{\pm}_\a
\end{align}
$R$ and $\bar{R}$ remain unchanged.\\
\textbf{Relation between the  AdS and flat basis:}
\begin{align}
P_a&=\frac{J_a-\bar J_a}{l}\;,\quad \cJ_a=J_a+\bar J_a\;,\quad \cQ^{1\pm}_\a=\sqrt{\frac{2}{l}}Q^\pm_\a\;,\quad \cQ^{2\pm}_\a=\sqrt{-\frac{2}{l}}\bar Q^\pm_\a\;,
\nonumber\\
 \cR&=R-\bar R \;,\quad \cS=\frac{R+\bar R}{l}\;,\quad A^a=\o^a+\frac{1}{l}e^a
\;,\quad \bar A^a=\o^a-\frac{1}{l}e^a
\nonumber\\
\psi^{1\a}_\pm&=\sqrt{\frac{l}{2}}\psi^\a_\pm\;,\quad \psi^{2\a}_\pm=\sqrt{-\frac{l}{2}}\eta^\a_\pm\;,\quad \sigma=l\frac{\phi_R+\bar\phi_R}{2}\;,\quad \upsilon=\frac{\phi_R-\bar\phi_R}{2}
\nonumber\\
\theta^{1\a}_\pm&=\sqrt{\frac{l}{2}}\varepsilon^\a_\pm\;,\quad \theta^{2\a}_\pm=\sqrt{-\frac{l}{2}}\vartheta^\a_\pm\;,\quad \l_\cR=\frac{\l_R-\bar\l_R}{2}\;,\quad \l_\cS=l\frac{\l_R+\bar\l_R}{2} \nonumber \\
\end{align}
\textbf{Relation between the two flat basis:}
\begin{align}
M_n&=P_a U^a_n\;,\quad \cL_n=\cJ_a U^a_n\;,\quad q^{1\pm}_\a=\sqrt{2}Q^{1\pm}_\a\;,\quad q^{2\pm}_\a=\sqrt{2}Q^{2\pm}_{-\a}\;,\quad 
%\nonumber\\
%\Psi^{1\a}_\pm&=\sqrt{2}\psi^{1\a}_\pm\;,\quad \Psi^{2\a}_\pm=\sqrt{-2}\psi^{2 \a}_{\pm}\;,\quad \phi=2\tilde\phi\;,\quad \rho=2\tilde\rho
%\nonumber\\
%\z^{1\a}_\pm&=\frac{1}{\sqrt{2}}\theta^{1\a}_\pm\;,\quad \z^{2\a}_\pm=\frac{1}{\sqrt{-2}}\theta^{2\a}_\pm\;,\quad \l_\cR=\l_\cR\;,\quad \l_\cS=\l_\cS
\end{align}
$\cR$ and $\cS$ remain unchanged.\\
\textbf{Relation between flat and AdS basis :}
\begin{align}
M_n&=\frac{L_n+\bar L_{-n}}{l}\;,\quad \cL_n=L_n-\bar L_{-n}\;,\quad  q^{1\pm}_\a=\sqrt{\frac{2}{l}}\cQ^\pm_\a\;,\quad q^{2\pm}_\a=\sqrt{\frac{2}{l}}\bar\cQ^\pm_{-\a}\;
\nonumber\\
 \cR&=R-\bar R\;,\qquad\quad \cS=\frac{R+\bar R}{l}\;,\quad \psi^{A\a}_\pm=\frac{1}{\sqrt{2l}}\Psi^{1,\a}_\pm\;,\quad \bar\psi^{A-\a}_\pm=\frac{1}{\sqrt{2l}}\Psi^\a_\pm\;,
 \nonumber\\
\phi&=l(\phi_R^A-\bar\phi_R^A)\;,\quad \rho=\phi_R^A+\bar\phi_R^A \;, \quad \l_\cR=\frac{\l_R^A-\bar\l_R^A}{2}\;,\quad \l_\cS=l\frac{\l_R^A+\bar\l_R^A}{2}\;,
\nonumber\\
\U^n&=\frac{\chi^n-\bar\chi^{-n}}{2}\;,\quad \xi^n=l\frac{\chi^n+\bar\chi^{-n}}{2}\;,\quad \epsilon^\a_\pm=\sqrt{\frac{2}{l}}\z^{1\a}_\pm\;,\quad \bar\epsilon^{-\a}_\pm=\sqrt{\frac{2}{l}}\z^{2\a}_\pm
\end{align}

\end{document}